# Zero-bias conductance peak in Dirac semimetal-superconductor devices

W. Yu[1,#], Rafael Haenel[2,#], M.A. Rodriguez[1], S.R. Lee[1], F. Zhang[3], M. Franz[2], D. I. Pikulin[4,*], and W. Pan[1,5,*]

[1]Sandia National Laboratories, Albuquerque, New Mexico 87185, USA

[2]Department of Physics and Astronomy, University of British Columbia, Vancouver, BC, Canada V6T 1Z1

[3]Department of Physics, University of Texas at Dallas, Dallas, Texas 75080, USA

[4]Microsoft Quantum, Microsoft Station Q, University of California, Santa Barbara, California 93106, USA

[5]Sandia National Laboratories, Livermore, California 94551, USA

[*]dmpikuli@microsoft.com; wpan@sandia.gov

[#]these authors contributed equally: Wenlong Yu, Rafael Haenel

Majorana zero modes (MZMs), fundamental building blocks for realizing topological quantum computers, can appear at the interface between a superconductor and a topological material. One of the experimental signatures that has been widely pursued to confirm the existence of MZMs is the observation of a large, quantized zero-bias conductance peak (ZBCP) in the differential conductance measurements. In this Letter, we report observation of such a large ZBCP in junction structures of normal metal (titanium/gold Ti/Au) – Dirac semimetal (cadmium-arsenide $Cd_3As_2$) – conventional superconductor (aluminum Al), with a value close to four times that of the normal state conductance. Our detailed analyses suggest that this large ZBCP is most likely *not* caused by MZMs. We attribute the ZBCP, instead, to the existence of a supercurrent between two far-separated superconducting Al electrodes, which shows up as a zero-bias peak because of the circuitry and thermal fluctuations of the supercurrent phase, a mechanism conceived by Ivanchenko and Zil'berman more than 50 years ago [JETP **28**, 1272 (1969)]. Our results thus call for extreme caution when assigning the origin of a large ZBCP to MZMs in a multiterminal semiconductor or topological insulator/semimetal setup. We thus provide criteria for identifying when the ZBCP is definitely not caused by an MZM. Furthermore, we present several remarkable experimental results of a supercurrent effect occurring over unusually long distances and clean perfect Andreev reflection features.

Topological quantum computation (QC) [1,2] has emerged as a promising approach to QC due to its enhanced tolerance to errors which are induced by inevitable coupling to the environment. In this scheme, a quantum bit (qubit) would be constructed out of four Majorana quasiparticles (MQPs) [3], with a gate performed by moving one MQP around another, or braiding MQPs. Because the braiding is topological and nonlocal, Majorana-based qubits would be intrinsically robust against local decoherence sources and thereby enable fault-tolerant QC.

Until 2008, the search for MQPs in solid-state systems focused on the exotic 5/2 fractional quantum Hall state [4,5], for which the earlier theoretical framework of TQC was developed [2]. However, concrete evidence of such quantum Hall non-Abelian anyons has not been established. The advent of topological insulators has prompted new proposals for realizing Majorana quasiparticles via the superconducting proximity effect [6,7]. In these systems, the Majorana quasiparticles are predicted to appear as topological defects, namely Majorana zero modes (MZMs), at the superconductor-magnet interface or π-Josephson junction at the surface of a topological insulator. Soon after this pioneering theoretical work, it was realized that MZMs could also be realized in a semiconductor system in combination with a strong spin-orbit coupling, an external magnetic field, and the superconducting proximity effect [8-10]. One of the experimental signatures that has been widely pursued to confirm the existence of MZMs is the observation of a large, quantized in units of $2e^2/h$ zero-bias conductance peak (ZBCP) in the differential conductance measurements. ZBCPs have been observed in InSb and InAs nanowire systems [11-17] and ferromagnetic atom chains [18]. However, the origin of the observed ZBCPs in these experiments remains debated [19-22]. As such, more experiments are imperative to establish MZMs in other topological quantum material systems beyond semiconducting nanowires. In the following, we will present results in a Dirac semimetal [23,24] cadmium-arsenide ($Cd_3As_2$) that show a large, nearly quantized in units of normal state conductance ZBCP in junction structures of normal metal (titanium/gold Ti/Au) – $Cd_3As_2$ – conventional superconductor (aluminum Al). Our detailed analyses suggest that this large ZBCP is most likely not caused by MZMs. We attribute the ZBCP, instead, to the existence of a supercurrent between two far-separated superconducting Al electrodes, which shows up as a zero-bias peak because of the circuitry and thermal fluctuations of the supercurrent phase, a mechanism conceived by Ivanchenko and Zil'berman more than 50 years ago [25-27]. Our results thus call for extreme caution when assigning the origin of a large ZBCP to MZMs in a multiterminal semiconductor or topological insulator/semimetal setup. We thus provide criteria for identifying when the ZBCP is definitely not caused by an MZM.

We choose $Cd_3As_2$ in this study due to the following reasons. First, it has been shown that the proximitized Fermi arc surface states [24] in $Cd_3As_2$ can show properties of a topological superconductor [28-30]. The induced pairing in these topological superconducting surface states can give rise to topologically protected gapless Andreev bound states, i.e., Majorana zero modes. Second, to our knowledge, $Cd_3As_2$ is the only topological semimetals system where the evidence of Majorana flat bands [31] have been observed [28]. This provides high confidence that localized Majorana zero modes can also be realized in $Cd_3As_2$. Third, as will see below, the decoupling of the surface and bulk states in $Cd_3As_2$ enables the Josephson effect that gives rise to the large ZBCP over a distance much longer than the coherence length. Finally, $Cd_3As_2$ is air stable and less prone to the oxidization issue that has troubled other topological materials such as $ZrTe_5$ and $Bi_2Te_3$. This makes $Cd_3As_2$ promising for practical quantum computation application. To search for MZMs, we follow the methodology developed in the InAs and InSb nanowire research [11-17] and fabricate Ti/Au-$Cd_3As_2$-Al junction structures. The mechanical exfoliation method is used to obtain the most flat and shiny Cd$_3$As$_2$ thin flakes for device fabrication from the initial ingot materials [28]. The Cd$_3$As$_2$ polycrystalline ingots used in this study are the same as those in Ref. [28]. The thickness of Cd$_3$As$_2$ flakes is about 200 nm. To fabricate Ti/Au-Cd$_3$As$_2$-Al junctions, we deposit thin flakes of Cd$_3$As$_2$ on a Si/SiO$_2$ substrate followed by two-step electron beam lithography to define Ti/Au and Al electrodes, respectively. The thickness of the electrodes is 10 nm/200 nm for Ti/Au and 300 nm for Al. A scanning electron microscopy image of one device is shown in Fig. 1a. A total of five junctions are fabricated in this device and they share the same Ti/Au electrode. The distance between two nearby junctions is ~ 1.5 μm in the device shown in Fig. 1a, but is larger in other devices. In our experiment, only the right most junction is measured, and the four remaining Al contact leads are floating. The distance between the Ti/Au and Al electrodes is about 80 nm. Fig. 1b shows the temperature ($T$) dependence of the resistance of this tunneling junction device. The superconducting transition temperature of Al electrode is at ~ 1.2 K (not shown). The sharp drop at $T$ ~ 0.7 K is due to the onset of the proximity effect induced bulk superconductivity in Cd$_3$As$_2$.

Metal-superconductor junction device structures have been extensively utilized in the past for studying superconductivity [32]. Similar structures have also been used in studying excess conductance in two-dimensional electron gas in III-V semiconductors [33,34]. When such a device is biased above the superconducting gap, the device is in the normal state, the current($I$) – voltage($V$) relationship is ohmic, and a constant differential conductance (d$I$/d$V$) is measured. When the device bias is reduced and the incident electron from the normal metal is within the superconducting energy gap, the electron can form a Cooper pair with another electron in the metal, and this Cooper pair then tunnels into the superconductor. Due to this process, a hole with the opposite spin and velocity but equal absolute momentum is reflected to the metal. This process, known as Andreev reflection (AR), is schematically depicted in Figure 1c. If the interface quality is high, the conductance within superconducting energy gap becomes twice the normal conductance [32], and this is commonly dubbed perfect AR (PAR). Figure 1d shows a clear example of PAR observed in our device at $T = 0.39$K, where the normalized differential conductance d$I$/d$V$ is plotted as a function of d.c. voltage bias ($V_{dc}$) across the junction, and normalization is defined relative to the normal state conductance, $G_N \approx 0.0128$ $\Omega^{-1}$. At high $|V_{dc}|$, the device is in the normal state and the conductance is constant, as expected. As $|V_{dc}|$ decreases, moving into the superconducting gap, the conductance assumes a value of 2$G_N$. One can argue that the observed PAR is a signature of Klein tunneling in the $Cd_3As_2$ surface state [35]. The observation of this PAR also demonstrates 1) the formation of a hard gap induced by the proximity effect and 2) the absence of backscattering in the normal region.

As the junction is further cooled down, a large ZBCP is observed. The normalized conductance at the base temperature of $T = 21$ mK and zero magnetic field is shown in Fig. 2a. Overall, the curve is very similar to that in Fig. 1d, except at $V_{dc} = 0$V, where a large, sharp peak is observed. Notably, the value of this peak is 4 times that of the normal conductance value; this near-integer increase most likely is not, in itself, of fundamental significance, as modeling and discussion below will ultimately demonstrate. To further examine the origin of this ZBCP, its temperature dependence is mapped out. Figure 2b shows the d$I$/d$V$ curves as a function of $V_{dc}$ at various temperatures. In Fig. 2c, the value of ZBCP is plotted versus $T$. At low temperatures, the amplitude is almost constant around 4 $G_N$. It then starts decreasing at $T > 0.1$K and eventually disappears around $T \sim 0.4$K. Lorentzian fitting to each temperature curve is performed, from which the full width at half maximum (FWHM) of the ZBCP is obtained. In Figure 2d, we show the resulting temperature dependence of the FWHM. It is nearly constant at $\sim 3$ μV in the low temperature regime, where the amplitude of ZBCP is also constant. the FWHM starts increasing above 0.1 K, where the ZBCP amplitude also starts decreasing. The FWHM increases sharply for $T > 0.3$ K.

Below, we try to explain the data by considering two alternative mechanisms for ZBCP formation: 1) the existence of MZMs at the interface between a superconductor and a quantum spin Hall insulator stack/Dirac semimetal [31,36-38] and 2) a supercurrent mechanism that may give rise to the observed large ZBCP [25]. Our detailed theoretical modeling will show that the latter origin is the one most likely to apply.

For the first mechanism, it is known that Fermi arc surface states of a Dirac semimetal can be viewed equivalently as a momentum-space stack of many identical two-dimensional quantum spin Hall insulator (QSHI) layers [38]. In other words, according to the theoretical work [38], in the presence of momentum conservation our junction structure could be viewed as a collection of many identical copies of Ti/Au-2D QSHI-Al junction devices. At low temperatures, MZMs can form at the interface between the QSHI edge channels and the superconductor [36,37], and in the presence of backscattering they contribute to a large, quantized in units of $2Ne^2/h$ ZBCP. Here $N$ is the number of channels in the junction.

However, our detailed examinations cast doubts over this mechanism. First, the width of ZBCP is too narrow, being even smaller than the measurement temperature (indicated by the red dash line in Fig. 2d), while the width of MZM ZBCPs cannot be smaller than 3.5 $K_BT$ [39]. This thus excludes a non-interacting electron explanation for the data. Second, magnetic backscattering is required to obtain localized MZMs that give rise to ZBCP. In the absence of magnetic scattering the weak antilocalization (WAL) effect in $Cd_3As_2$ should produce a zero-bias conductance dip [39]. This is inconsistent with the experimental observation.

Next, we consider the supercurrent mechanism and conduct an analysis in terms of an RSJ model [26] in the presence of thermal fluctuations, which was first solved by Ivanchenko and Zil'berman [25]. First, where does the supercurrent in our setup arise? We assume good ohmic contact between the Al and Ti/Au electrodes in each of the NS (normal metal-superconductor) junctions below the superconducting gap. Then two nearby Al electrodes, if they are phase coherent, can support supercurrents and additional current path (as shown by the green arrows in Fig. 1a) besides the direct NS one (as shown by the red arrow in Fig. 1a) is available. Thermal fluctuations and circuitry move the maximum of the supercurrent from the zero bias. In this case the supercurrent peak in $I$-$V$ characteristics shows up a ZBCP in $dI/dV$. The large distance between the superconducting electrodes suppresses the supercurrent making the observed amplitude of ZBCP comparable to $G_N$.

In the following, we use the IZ formula [25] to quantitatively fit our measured differential conductance curves and extract parameters of our system. For this purpose, we first obtain $I$-$V$ curves by integrating the differential conductance. Fig. 3A shows the result at $T = 21$ mK. The fitting range is limited within the Andreev reflection regime. Next, the linear contribution due to AR within the superconducting gap is subtracted from the $I$-$V$ curve as it is due to a parallel conductance channel. The obtained data represents the contribution from the supercurrent between two Al electrodes. The circuit diagram in the lower inset of Fig. 3A is an effective description for the experimental setup. Here, $V_B$ is the voltage source, $V$ the voltage that is plotted on the x-axis in Fig. 3A, $R_1$ the junction resistance, and $R_2$ the lump resistance in the measurement circuit. We neglect the capacitance *C* of the junction in the fitting. This is justified because of a large separation of Al electrodes.

The equations for the *I-V* curve fitting are then given as follows [25]:

$$I(V) = I_0 Im\left[\frac{I_{1-2i\beta(V+R_2I)\hbar/(2e(R_1+R_2))}\left(\beta\frac{\hbar}{2e}I_0\frac{\Delta(T)}{\Delta(0)}\right)}{I_{-2i\beta(V+R_2I)\hbar/(2e(R_1+R_2))}\left(\beta\frac{\hbar}{2e}I_0\frac{\Delta(T)}{\Delta(0)}\right)}\right] \quad (1)$$

$$\frac{\Delta(T)}{\Delta(0)} = 1 - \sqrt{\frac{2\pi kT}{\Delta(0)}}\exp\left(-\frac{\Delta(0)}{kT}\right) \quad (2)$$

Here, $I_n(x)$ is the modified Bessel function of complex order, $\Delta(0) = 50$ μeV is the topological superconductivity gap at $T = 0$, and $I_0$ the total critical current of the supercurrent quanta. The BCS temperature dependence of $\Delta(T)/\Delta(0)$ is the first order expansion around $T = 0$. In Figs. 3B, we show the fitting at four selected temperatures (fitting at more temperatures can be found in the SM [39]). It agrees well with the experimental data at low temperatures, e.g., at $T = 21, 124, 203$ mK. Fitting fails at higher temperatures, for example at $T = 350$ mK, where the temperature becomes comparable to the induced superconducting gap and quasiparticle contributions to current likely disturb the fitting. We note here that fitting using the phenomenological temperature dependence of *Δ(T)/Δ(0) = tanh (1.74(T_c/T-1)^{1/2})* yields the same conclusion. In Fig. 3C, the maximal current extracted from the fitting at each temperature is plotted.

The good agreement between the theoretical fitting and experimental data strongly supports that the ZBCP is due to the supercurrent between two separated Al electrodes. Moreover, because of this supercurrent origin, the FWHM width of ZBCP can be smaller than the measurement temperature, as the width of the ZBCP is determined by the circuit parameters, not intrinsic temperature broadening. This explains the puzzle seen in Fig. 2d.

The parameters obtained from the above fitting are $I_0 \approx 35$ nA, $R_1 \approx 60\ \Omega$, and $R_2 \approx 90\ \Omega$. $R_1$ is close to the normal-state resistance across the Ti/Au-Al junction, which is about 78 $\Omega$. Based on these fitting parameters, we can estimate the number of spin-resolved supercurrent quanta in our device. First, we calculate the critical current $\Delta I_0$ of a single supercurrent quantum. It is given by $\Delta I_0 = \Delta(0) \times \pi/e \times G = 6.1$ nA [40]. Here, $G = e^2/h$ for the helical (effectively spin-polarized) topological surface states of $Cd_3As_2$. Consequently, the total number of the supercurrent quanta is 34.8/6.1 ~ 6.

Notably, the supercurrent mechanism also explains the peculiar magnetic ($B$) field dependence of the ZBCP we observed. In Figure 4, the measured d$I$/d$V$ is plotted as a function of magnetic field and $V_{dc}$, respectively. The ZBCP in our device is extremely sensitive to the magnetic fields and disappears at a magnetic field merely larger than 1.3 mT. Figure 4b shows the extracted experimental values of ZBCP at different temperatures demonstrating a clear oscillating pattern. For example, at the base $T$ of 21 mK, the amplitude decreases from $4G_N$ at $B = 0$ T to ~ $2G_N$ at 0.5 mT. It then increases to $3G_N$ at 0.75 mT, before it drops to zero again at higher magnetic fields. Since the critical magnet field of the superconducting Al electrodes is ~ 40 mT, much larger than 1.3 mT, the disappearance cannot be due to the loss of the superconductivity in Al. Furthermore, this small period cannot be caused by the magnetic flux across the Ti/Au-Al junction area. In fact, considering the area of the junction is $S \sim 4 \times 10^{-7} m^2$, magnetic flux quantum is inserted into the junction at $B = \frac{\varphi}{S} = 52$ mT, much larger than the observed 0.5 mT. Here $\varphi = 2.067 \times 10^{-15}$ Wb is magnetic flux quantum. Thus, interference across this junction can be excluded as the source of the ZBCP oscillations. On the other hand, under the supercurrent model, the effective junction area should be the one between the two adjacent junction devices, which is about 4 μm². Consequently, the Fraunhofer period is ~ 0.5 mT, consistent with the experimental observation.

Finally, one remark is in order before we conclude. The IZ model was conventionally used for short Josephson junctions. In our case, however, it still applies to a device where the two Al electrodes are separated by ~ 10 μm, as shown in Fig. 5a. In this device, the normal metal contacts are made of Ni/Au. The differential conductance between the Ni/Au and Al electrodes for the middle junction is shown in Fig. 5b. Overall, it is similar to that in the first device already discussed, with both exhibiting a perfect AR and a large ZBCP. Applying the IZ fitting yields again a good agreement, as shown in Fig. 5c for the 98 mK data and in Fig. 5d for the extracted $I_0$. However, unlike the first specimen where the separation between two adjacent junctions is comparable to the Al superconducting coherence length ($\xi_{Al}$ ~ 1μm), $Cd_3As_2$ phase coherence length ($l_\phi$ ~ 0.6 μm) [40], and/or $Cd_3As_2$ thermal length $L_T$ (at $T$ = 20 mK, $L_T$ ~ 1.1 μm is estimated using a diffusion constant of $D$ ~ 0.02 m²/s [41]), the separation of Al electrodes in this device is now ~10 μm, much longer than $\xi_{Al}$ and $l_\phi$. Thus, it is totally surprising that the supercurrent appears to still exist over such a large distance. In the following, we discuss three possibilities to resolve this apparent difficulty in our interpretation. First, $l_\phi$ entering the WAL measurements [42] is different from $l_\phi$ relevant for the Aharonov-Bohm-like interference necessary for the Fraunhofer patter [43]. Second, the supercurrent is proportional to the transmission (to the power of 2) through the channels connecting the superconductors. In our device with large Al electrodes, there can exist a tail of the channel transmission distribution responsible for the supercurrent. Third, it is possible that existence of supercurrent is determined by the electron pairing coherence length, which might be very long in the topological surface states in $Cd_3As_2$ due to the suppressed backscattering. Finally, the Josephson effect over such a long distance seems to indicate relative decoupling of the surface and bulk states which can be used for potential applications of the material.

In summary, we report observation of a large ZBCP in tunnel junctions made of normal metal (Ti/Au) – Dirac semimetal ($Cd_3As_2$) – conventional superconductor (Al). Our detailed analyses suggest that this large ZBCP probably is due to the existence of a supercurrent between two far-separated superconducting Al electrodes. Our results thus call for extreme caution when assigning a large ZBCP to the MZM origin, especially when the width of the ZBCP is below $3.5K_BT$.

The work at Sandia National Labs was supported by a Laboratory Directed Research and Development project. Device fabrication was performed at the Center for Integrated Nanotechnologies, an Office of Science User Facility operated for the U.S. Department of Energy (DOE) Office of Science. Sandia National Laboratories is a multimission laboratory managed and operated by National Technology & Engineering Solutions of Sandia, LLC, a wholly owned subsidiary of Honeywell International Inc., for the U.S. Department of Energy's National Nuclear Security Administration under contract DE-NA0003525. This paper describes objective technical results and analysis. Any subjective views or opinions that might be expressed in the paper do not necessarily represent the views of the U.S. Department of Energy or the United States Government. Work at UT Dallas was supported by ARO under Grant No. W911NF-18-1-0416 and NSF under Grant No. DMR-1921581. Work at UBC was supported by NSERC and CIFAR.

**References**

[1] A. Yu. Kitaev, Annals of Physics **303**, 2 (2003).

[2] C. Nayak, S.H. Simon, A. Stern, M. Freedman, and S. Das Sarma, Rev. Mod. Phys. **80**, 1083 (2008).

[3] S.R. Elliott and M. Franz, Rev. Mod. Phys. **87**, 137-163 (2015).

[4] R. Willett, J.P. Eisenstein, H.L. Störmer, D.C. Tsui, A.C. Gossard, and J.H. English, Phys. Rev. Lett. **59**, 1776 (1987).

[5] W. Pan, J.-S. Xia, V. Shvarts, D.E. Adams, H.L. Stormer, D.C. Tsui, L.N. Pfeiffer, K.W. Baldwin, and K.W. West, Phys. Rev. Lett. **83**, 3530 (1999).

[6] L. Fu and C.L. Kane, Phys. Rev. Lett. **100**, 096407 (2008).

[7] M. Franz, Nature Nanotechnology **8**, 149 (2013).

[8] R.M. Lutchyn, J.D. Sau, and S. Das Sarma, Phys. Rev. Lett. **105**, 077001 (2010).

[9] Y. Oreg, G. Refael, and F. von Oppen, Phys. Rev. Lett. **105**, 177002 (2010).

[10] J. Alicea, Physical Review B **81**, 125318 (2010).

[11] K. Mourik, K. Zuo, S.M. Frolov, S.R. Plissard, E.P.A.M. Bakkers, and L.P. Kouwenhoven, Science **336**, 1003 (2012).

[12] L.P. Rokhinson, X. Liu, and J.K. Furdyna, Nature Physics **8**, 795 (2012).

[13] A. Das, Y. Ronen, Y. Most, Y. Oreg, M. Heiblum, and H. Shtrikman, Nature Physics **8**, 887 (2012).

[14] M.T. Deng, C.L. Yu, G.Y. Huang, M. Larsson, P. Caroff, and H.Q. Xu, Nano Lett. **12**, 6414 (2012).

[15] A.D.K. Finck, D.J. Van Harlingen, P.K. Mohseni, K. Jung, and X. Li, Phys. Rev. Lett. **110**, 126406 (2013).

[16] M.T. Deng, S. Vaitiekenas, E.B. Hansen, J. Danon, M. Leijnse, K. Flensberg, J. Nygård, P. Krogstrup, and C.M. Marcus, Science **354**, 1557 (2016).

[17] H. Zhang, C.-X. Liu, S. Gazibegovic, D. Xu, J.A. Logan, W. Wang, N. van Loo, J.D.S. Bommer, M.W.A. de Moor, D. Car, R.L. Op het Veld, P.J. van Veldhoven, S. Koelling, M.A.Verheijen, M. Pendharkar, D.J. Pennachio, B. Shojaei, J.S. Lee, C.J. Palmstrøm, E.P.A.M. Bakkers, S. Das Sarma, and L.P. Kouwenhoven, Nature **556**, 74 (2018).

[18] S. Nadj-Perge, I.K. Drozdov, J. Li, H. Chen, S. Jeon, J. Seo, A.H. MacDonald, B.A. Bernevig, and A. Yazdani, Science **346**, 6209 (2014).

[19] K.T. Law, P.A. Lee, and T.K. Ng, Phys. Rev. Lett. **103**, 237001 (2009).

[20] C.-X. Liu, J.D. Sau, T.D. Stanescu, and S. Das Sarma, Phys. Rev. B **96**, 075161 (2017).

[21] C. Moore, C. Zeng, T.D. Stanescu, and S. Tewari, Phys. Rev. B **98**, 155314 (2018).

[22] O.A. Awoga, J. Cayao, and A.M. Black-Schaffer, Phys. Rev. Lett. **123**, 117001 (2019).

[23] S.M. Young, S. Zaheer, J.C.Y. Teo, C.L. Kane, E.J. Mele, and A.M. Rappe, Physical Review Letters **108**, 140405 (2012).

[24] Z. Wang, H. Weng, Q. Wu, X. Dai, and Z. Fang, Physical Review B **88**, 125427 (2013).

[25] Yu. M. Ivanchenko and L.A. Zil'berman, JETP **28**, 1272 (1969).

[26] Michael Tinkham, Introduction to superconductivity, Courier Corporation (2004).

[27] R. L. Kautz and John M. Martinis, Phys. Rev. B **43**, 9903 (1990).

[28] W. Yu, W. Pan, D.L. Medlin, M.A. Rodriguez, S.R. Lee, Z.-Q. Bao, and F. Zhang, Phys. Rev. Lett. **120**, 177704 (2018).

[29] A.-Q. Wang, C.-Z. Li, C. Li, Z.-M. Liao, A. Brinkman, and D.-P. Yu, Phys. Rev. Lett. **121**, 237701 (2018).

[30] F. Zhang and W. Pan, Nature Materials **17**, 851 (2018).

[31] A. Chen, D.I. Pikulin, and M. Franz, Phys. Rev. B **95**, 174505 (2017).

[32] G.E. Blonder, M. Tinkham, and T.M. Klapwijk, Rev. Rev. B **25**, 4515 (1982).

[33] A. Kastalsky, A. W. Kleinsasser, L. H. Greene, R. Bhat, F. P. Milliken, and J. P. Harbison, Phys. Rev. Lett. **67**, 3026 (1991).

[34] F. Rahman, T. J. Thornton, R. Huber, L. F. Cohen, W. T. Yuen, and R. A. Stradling, Phys. Rev. B **54**, 14026 (1996).

[35] S. Lee, V. Stanev, X. Zhang, D. Stasak, J. Flowers, J.S. Higgins, S. Dai, T. Blum, X. Pan, V.M. Yakovenko, J. Paglione, R.L. Greene, V. Galitski, and I. Takeuchi, Nature **570**, 344–348 (2019).

[36] L. Fu and C.L. Kane, Phys. Rev. B 79, 161408(R) (2009).

[37] H. Wang, H.C. Wang, H.W. Liu, H. Lu, W.H. Yang, S. Jia, X.-J. Liu, X.C. Xie, J. Wei, and J. Wang, Nat. Mater. **15**, 38 (2016).

[38] A. Chen and M. Franz, Phys. Rev. B **93**, 201105(R) (2016).

[39] See Supplemental Materials for detailed information.

[40] N.M. Chtchelkatchev, G.B. Lesovik, and G. Blatter, Phys. Rev. B **62**, 3559 (2000).

[41] M. N. Chernodub and M. A. Zubkov, Phys. Rev. B **95**, 115410 (2017).

[42] B. Zhao, P. Cheng, H. Pan, S. Zhang, B. Wang, G. Wang, F.X. Xiu, and F.Q. Song, Scientific Reports **6**, 22377 (2016).

[43] T. Ludwig and A.D. Mirlin, Phys. Rev. B **69**, 193306 (2004).

**Figures and Figure Captions**

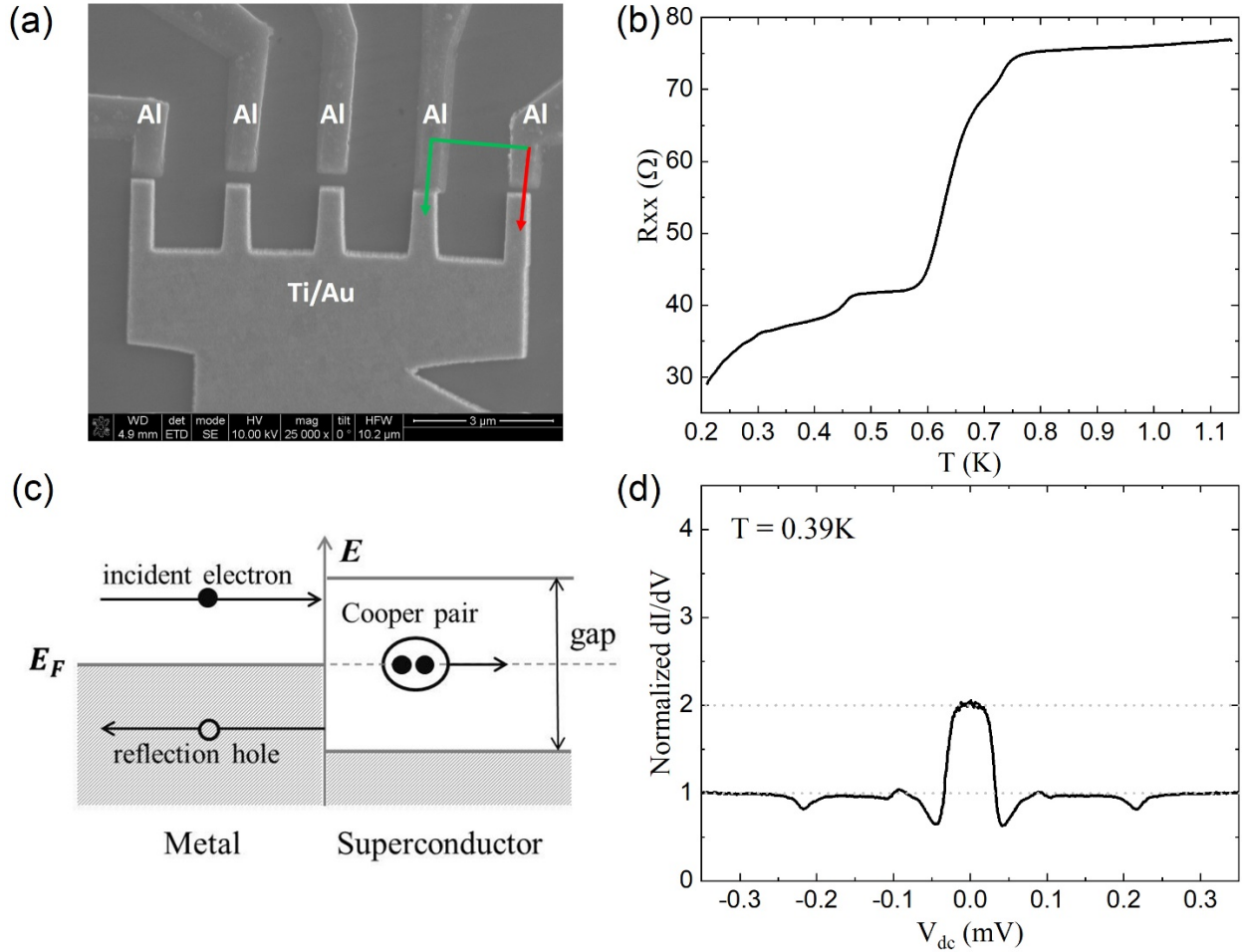

**Figure 1**: (a) An SEM picture of the device. The Ti/Au and Al electrodes are deposited on the Cd$_3$As$_2$ thin flake. (b) Temperature dependence of the junction resistance. (c) A schematic diagram showing the Andreev reflection. (d) The normalized differential conductance d$I$/d$V$ versus the d.c. bias across the junction ($V_{dc}$) measured in our device at $T = 0.39$K at zero magnetic field.

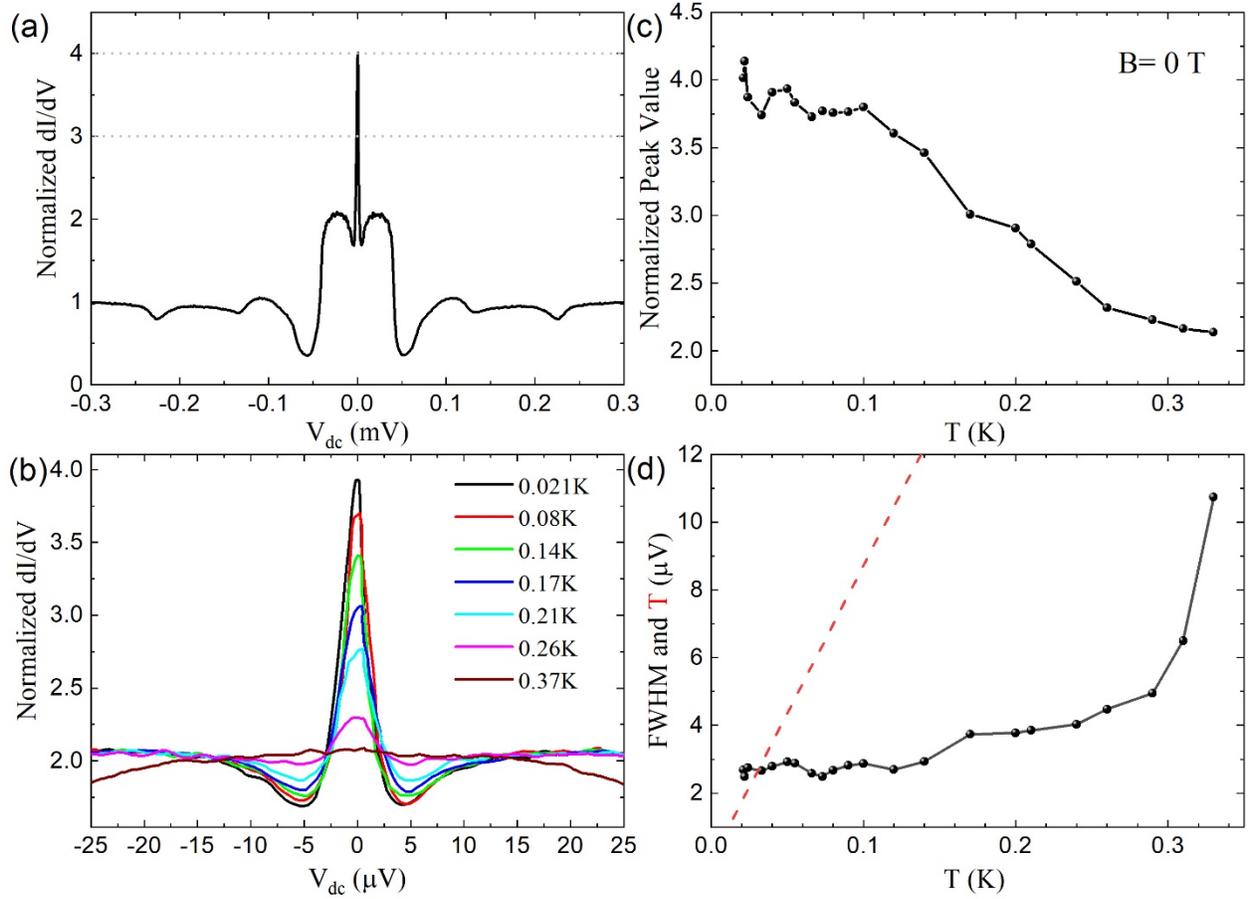

**Figure 2**: (a) d$I$/d$V$ at base temperature $T$ = 21 mK at zero magnetic field. A zero-bias conductance peak (ZBCP) is observed. The value of ZBCP is 4 times of the normal state conductance $G_N$. (b) d$I$/d$V$ at various temperatures. (c) The peak value as a function of temperature. The value is nearly quantized at low temperatures and starts to decrease at $T$ is higher than 100 mK. (d) The full width at half magnitude of ZBCP as a function of temperature. The red dashed line indicates the junction temperature.

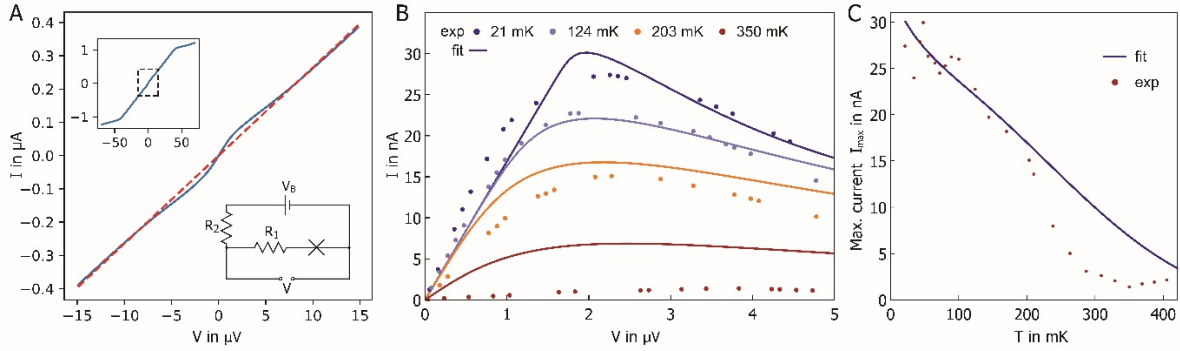

**Figure 3**: (A) the *I-V* curve at T = 21 mK, obtained by integrating the corresponding d*I*/d*V* curve. Here only the data in the AR regime is shown. The red dash line represents the contribution from AR. The up inset shows the whole I-V curve. The lower inset shows the effective circuit for the experimental setup. (B) IZ fitting at four selected temperatures. Dots are experimental data, line the IZ fitting. (C) the extracted maximal current at each temperature. The solid line is the theoretical fitting.

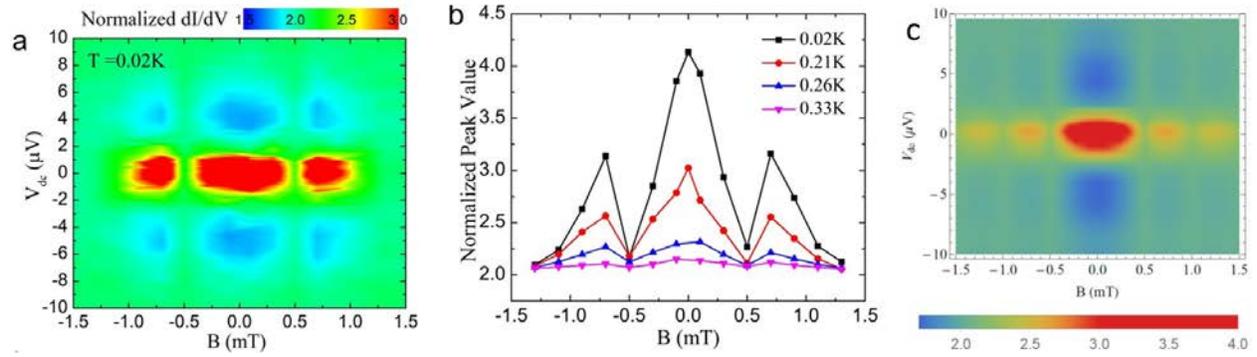

**Figure 4**: **(**a) 2D color plot of d*I*/d*V* as a function of magnetic field and $V_{dc}$, respectively. ZBCP disappears at the magnetic field of $B \sim 0.5$ mT and for $B > 1.3$ mT. (b) The extracted peak value as a function of magnetic field at various temperatures. an oscillatory behavior is observed for the ZBCP value. (c) Theoretical simulation of the magnetic field dependence with a Fraunhofer pattern based on the experimental data at zero magnetic field. Good agreement between the experimental result (a) and theoretical simulation (c) is seen.

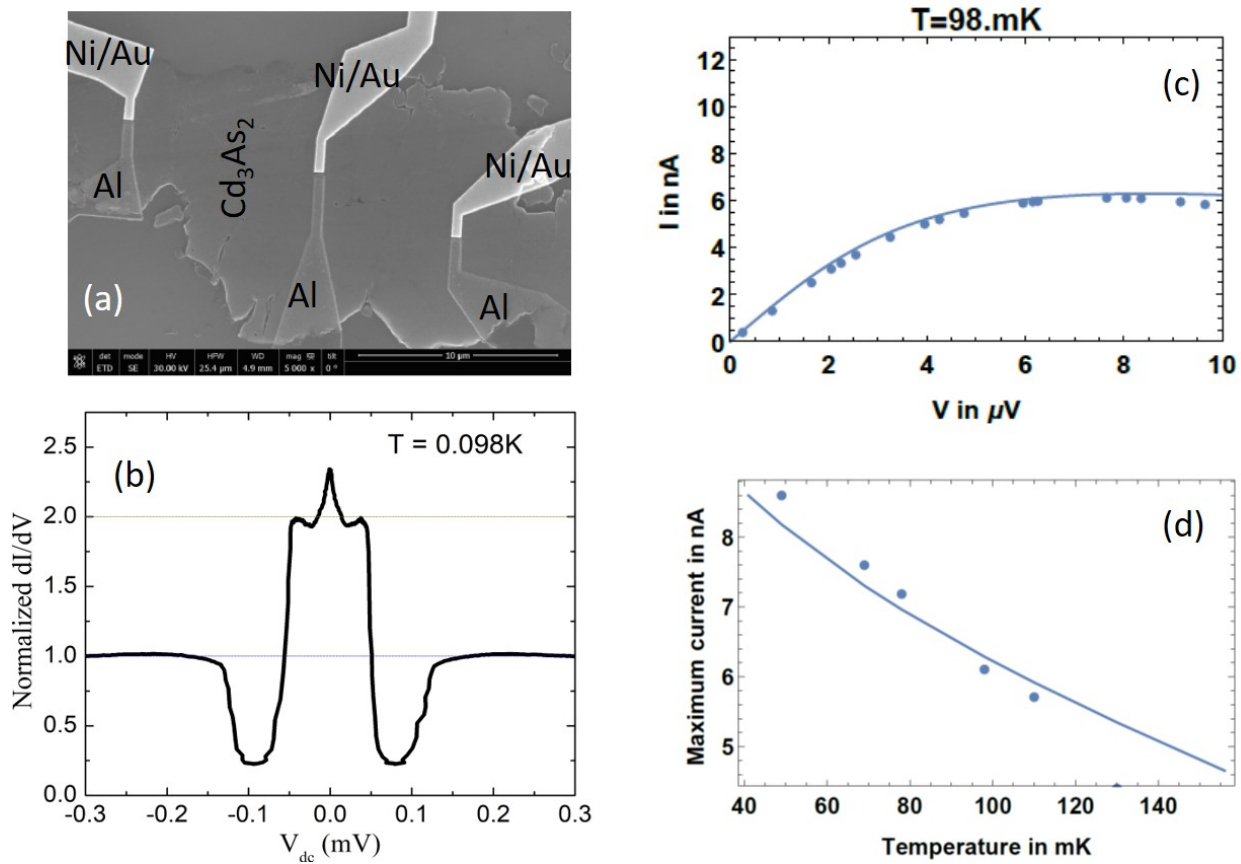

**Figure 5**: (a) SEM image of another device where the tunnel junctions are separated by ~ 10 μm. The irregular shaped Cd$_3$As$_2$ thin flake is visible. (b) large ZBCP measured in the middle junction at $T$ = 98 mK. (c) IZ fitting to the $T$ = 98 mK data. (d) $I_0$ as a function of temperature. The solid line is the theoretical fitting.

# SUPPLEMENTAL MATERIAL

# Zero Bias Conductance Peak in Dirac Semimetal-Superconductor Devices

W. Yu[1,#], Rafael Haenel[2,#], M.A. Rodriguez[1], S.R. Lee[1], F. Zhang[3], M. Franz[2], D. I. Pikulin[4,*], and W. Pan[1,5,*]

[1]*Sandia National Laboratories, Albuquerque, New Mexico 87185, USA*

[2]*Department of Physics and Astronomy, University of British Columbia, Vancouver, BC, Canada V6T 1Z1*

[3]*Department of Physics, University of Texas at Dallas, Dallas, Texas 75080, USA*

[4]*Microsoft Quantum, Microsoft Station Q, University of California, Santa Barbara, California 93106, USA*

[5]*Sandia National Laboratories, Livermore, California 94551, USA*

To explore a possible Majorana origin of the experimentally observed zero bias conductance peak (ZBCP) we conducted extensive tight-binding simulations. To this end, we adopt the low energy model for $Cd_2As_3$ [1, 2]

$$H_0 = \epsilon_k + \begin{pmatrix} M_k & A\,k_- & 0 & 0 \\ A\,k_+ & -M_k & 0 & 0 \\ 0 & 0 & -M_k & -A\,k_- \\ 0 & 0 & -A\,k_+ & M_k \end{pmatrix}, \quad (1)$$

where, $k_\pm = k_x \pm i k_y$, $\epsilon_k = C_0 + C_1 k_z^2 + C_2(k_x^2 + k_y^2)$ and $M_k = M_0 + M_1 k_z^2 + M_2(k_x^2 + k_y^2)$. Model parameters are obtained from [2]. We further include small perturbations that couple the two blocks of Hamiltonian $H_0$ and are consistent with symmetry class DIII. The spectrum

$$E(\boldsymbol{k}) = \epsilon_k \pm \sqrt{M_k^2 + A^2(k_x^2 + k_y^2)}$$

has two doubly degenerate Dirac nodes at $(0,0,\pm Q)$ with $Q = \sqrt{\left|\frac{M_0}{M_1}\right|}$. Let us treat the momentum variable $k_z$ as a parameter. Then, for $|k_z| < Q$, we can interpret

$$H_{|k_z|<Q}(k_x, k_y) = H(\mathbf{k}) \tag{2}$$

as a Quantum Spin Hall Insulator (QSHI) in two spatial dimensions $x$ and $y$. When the edge of such a QSHI is in proximity to an s-wave superconductor, we expect a Majorana zero mode (MZM) at each boundary of the superconducting region [3]. This is in direct analogy to the end states of a Kitaev wire [4]. In our case, however, the two MZMs are coupled via the metallic surface states that extend along the unproximitized edge of the 2D QSHI. This hybridizes the MZMs away from zero energy. To confine the Majorana modes to zero energy and localize them at the SC boundary, we need to gap out the metallic surface states. This is most conveniently achieved by a time-reversal symmetry (TRS) breaking perturbation.

Explicitly, we regularize Eq. (1) on a cubic lattice with lattice constant $a = 20$Å and add a superconducting profile $\Delta(y)$ as well as a TRS breaking Zeeman profile $J(x)$ as depicted in the top panel of Fig. S1. $N$ and $S$ denote semi-infinite normal and superconducting leads of width $\frac{L_z}{a} = 1080$ and height $\frac{L_x}{a} = 60$, respectively. In summary, the full model Hamiltonian is

$$H = \begin{pmatrix} H_0 - \mu(\mathbf{x}) + J_x(\mathbf{x}) & \Delta(\mathbf{x}) \\ \Delta^*(\mathbf{x}) & -(H_0 - \mu(\mathbf{x})) + J(\mathbf{x}) \end{pmatrix},$$

with

$$J(\mathbf{x}) = B(\mathbf{x}) \begin{pmatrix} 0 & 0 & 0 & -1 \\ 0 & 0 & -\frac{1}{2} & 0 \\ 0 & -\frac{1}{2} & 0 & 0 \\ -1 & 0 & 0 & 0 \end{pmatrix}$$

and the chemical potential $\mu$ is tuned into the Dirac point.

We use Kwant simulation libraries [5] to compute the differential conductance $G$ between the normal lead $N$ and the superconducting lead $S$. The result is shown in the bottom panel of Fig.

S1. The normal state conductance $G_N$ outside of the superconducting gap (gray shadowed area) is mostly due to transmission of the QSH edge mode at the bottom $yz$-surface. The conductance value roughly doubles for an applied bias within the superoconducting gap $\Delta\backslash e = 100\mu V$ as a result of perfect Andrev reflection.

At zero bias we observe the hallmark of the MZMs localized at the phase boundary between magnetic region (red) and SC lead (blue): a sharp zero bias conductance peak (ZBCP). Each surface state in the top $yz$-plane contributes a $\frac{2e^2}{h}$ Lorentzian peak to the ZBCP, yielding a total hight of the ZBCP of approximately $4G_N$. The width of each Lorentzian is proportional to the coupling strength to the Majorana mode. Close to the Dirac points $(0,0,\pm Q)$ the Majorana wavefunctions are less localized and couple strongly to the leads. Here, the Lorentzian width is large. This adds a long tail to the ZBCP and even contributes to the conductance outside of the superconducting gap. The latter fact explains the deviation of the ZBCP height from the expected value $4G_N$.

At finite temerature $T$ sharp features of the conductance are broadened. Most notably, the FWHM of the ZBCP is always wider than $3.5k_BT$ as highlighted by the color-shaded areas in Fig. S1. For our parameters at about $200mK$ the ZBCP is completely washed out. We emphasize that this is in disagreement with the experimental observations where the FWHM of the ZBCP stays constant over a temperature range from $21mK$ to $200mK$, and is below $3.5k_BT$ for much of the region. This is the main argument against a Majorana origin of the experimentally observed ZBCP. We note that the width below $3.5k_BT$ excludes *any* non-interacting mechanism as the origin of the ZBCP.

The existance of a MZM ZBCP in above simulations relies on the presence of a TRS breaking magnetic field that decouples the two Majorana modes at the boundary of the superconducting region. Notably, no such TRS breaking perturbation seems present in our experiment. Naively, an alternative way to gap out the Dirac semimetal surface is realized by chemical potential disorder coupling the different QSHI systems corresponding to the different momenta slices in Eq. (2). This approach, however, relies on fine-tuning of parameters. First, disorder needs to be strong, since backscattering between surface modes of opposite momenta is forbiddenby TRS.

On the other hand, disorder has to be weak enough to allow coupling of the normal lead to the localized MZMs. Despite extensive numerical simulations we were unable to identify a parameter regime that supports localized MZMs in the presence of TRS, thus rejecting the naïve hypothesis. The results of an exemplary quantum transport simulation are presented in Fig. S2. Here, a normal lead $N$ with six propagating modes and a superconducting lead $S$ connect to the scattering region $W$ with dimensions $L_x = L_x = 15a$ and $L_y = 800a$. The scattering region hosts chemical potential disorder generated from the box distribution $\left[-\frac{W}{2}, \frac{W}{2}\right]$. Instead of a ZBCP, we observe a zero bias conductance minimum resulting from Weak Localization (WL) [6]. These results are inconsistent with the experiment.

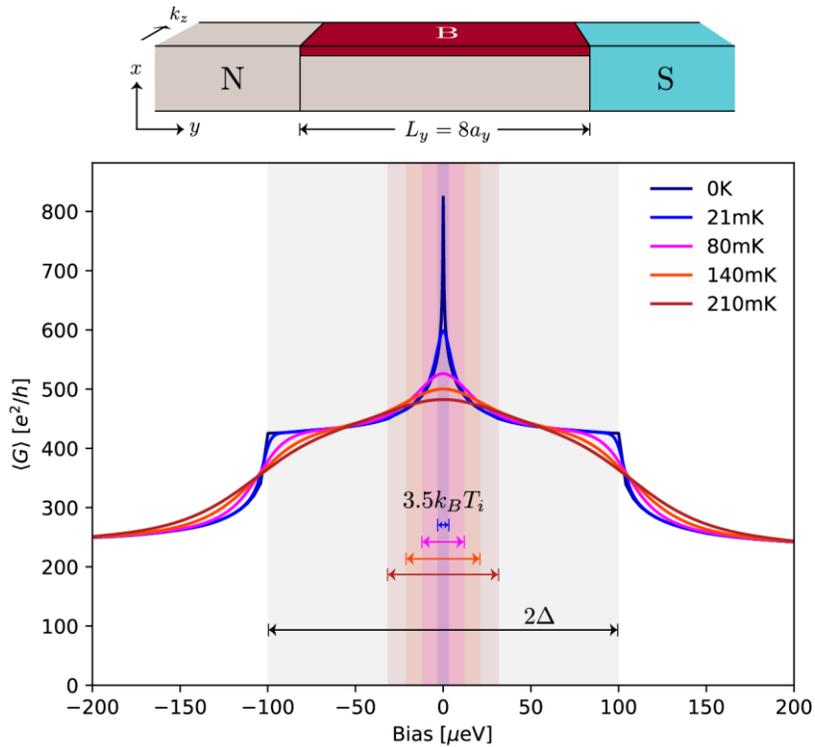

*Figure S 1 Quantum-transport simulation of the geometry depicted in the top panel. Perfect Andreev reflection of edge modes in the bottom yz-surface gives rise to a conductance plateau of twice the normal state conductance $2G_N$ within the superconducting gap $\Delta = 100\mu eV$. Tunneling of top yz-ssurface states into the MZMs adds a zero bias peak of height $2G_N$ to the conductance.*

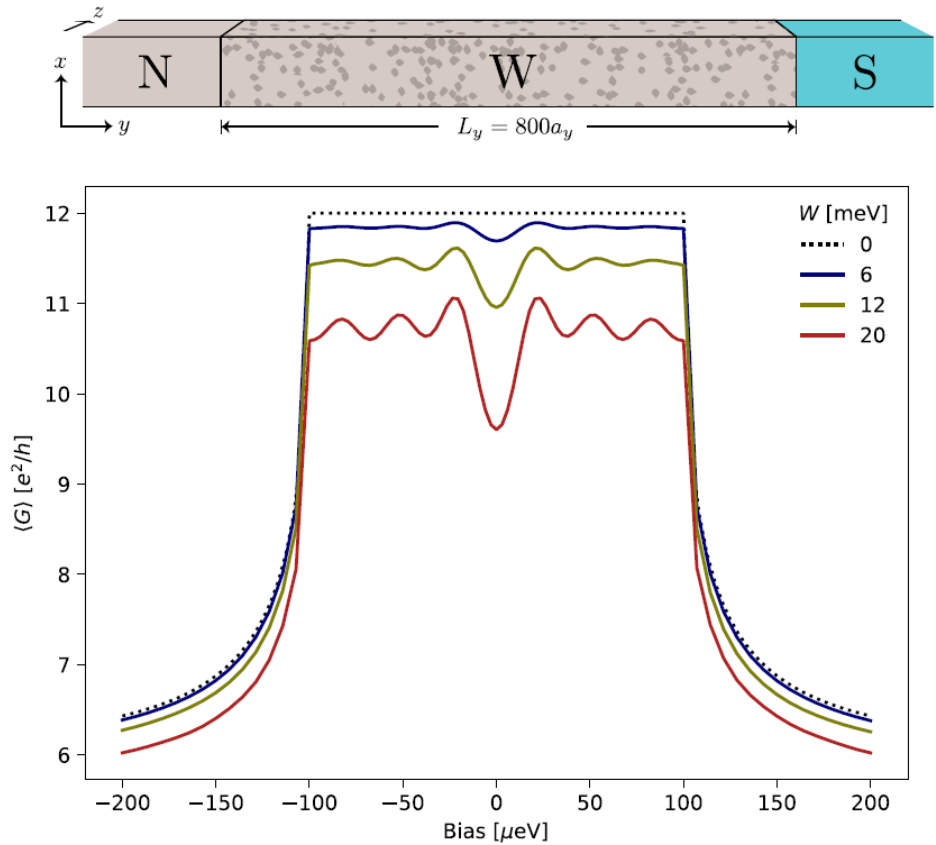

*Figure S 2 Zero bias conductance minimum from Weak Localization. The scattering region W has dimension $L_x = L_z = 15a$ and $L_y = 800a$. Disorder is simulated by a random chemical potential drawn from the box distribution $\left[-\frac{W}{2}, \frac{W}{2}\right]$ and averaged over 40 realizations.*

[1] Cano, J., Bradlyn, B., Wang, Z., Hirschberger, M., Ong, N. P., & Bernevig, B. A. (2017). Chiral anomaly factory: Creating Weyl fermions with a magnetic field. *Physical Review B*, *95*(16), 161306. https://doi.org/10.1103/PhysRevB.95.161306

[2] Wang, Z., Weng, H., Wu, Q., Dai, X., & Fang, Z. (2013). Three-dimensional Dirac semimetal and quantum transport in Cd 3As2. *Physical Review B - Condensed Matter and Materials Physics*, *88*(12), 125427. https://doi.org/10.1103/PhysRevB.88.125427

[3] Fu, L., & Kane, C. L. (2008). Superconducting proximity effect and majorana fermions at the surface of a topological insulator. *Physical Review Letters*, *100*(9), 096407. https://doi.org/10.1103/PhysRevLett.100.096407

[4] Kitaev, A. Y. (2001). Unpaired Majorana fermions in quantum wires. *Physics-Uspekhi*, *44*(10S), 131. https://doi.org/10.1070/1063-7869/44/10S/S29

[5] Groth, C. W., Wimmer, M., Akhmerov, A. R., & Waintal, X. (2014). Kwant: A software package for quantum transport. *New Journal of Physics*, *16*. https://doi.org/10.1088/1367-2630/16/6/063065

[6] Pikulin, D. I., Dahlhaus, J. P., Wimmer, M., Schomerus, H., & Beenakker, C. W. J. (2012). A zero-voltage conductance peak from weak antilocalization in a Majorana nanowire. *New Journal of Physics*, *14*. https://doi.org/10.1088/1367-2630/14/12/125011

# The I-V curve fitting at other temperatures

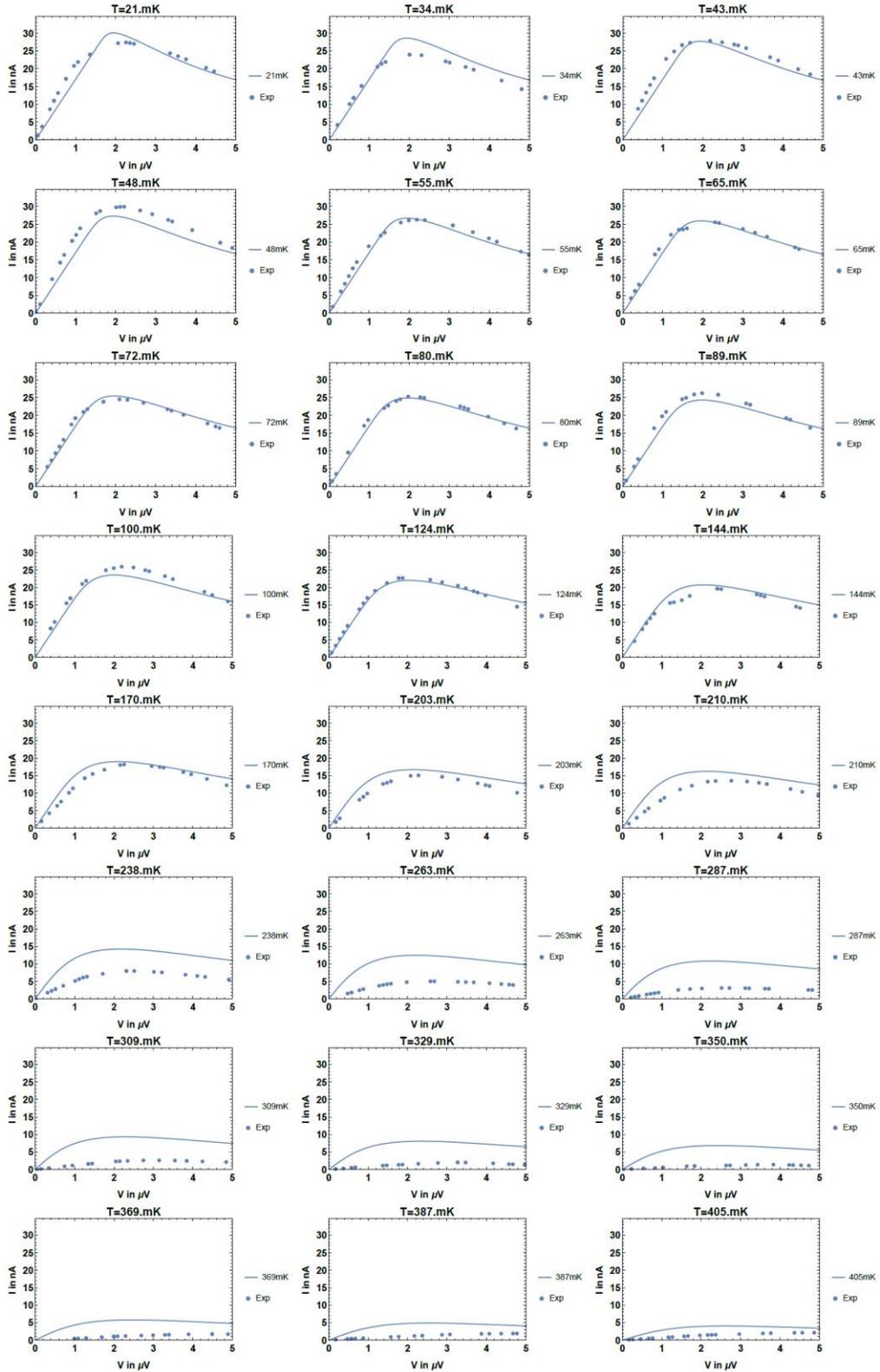

**Data sets in Fig. 3**

dIdV21mK.dat

| A | B |
|---|---|
| 1.826E-4 | 0.01213 |
| 1.819E-4 | 0.01209 |
| 1.807E-4 | 0.01212 |
| 1.801E-4 | 0.01214 |
| 1.796E-4 | 0.0121 |
| 1.785E-4 | 0.01208 |
| 1.787E-4 | 0.01209 |
| 1.778E-4 | 0.01201 |
| 1.77E-4 | 0.01206 |
| 1.757E-4 | 0.01198 |
| 1.757E-4 | 0.01203 |
| 1.743E-4 | 0.01211 |
| 1.739E-4 | 0.01216 |
| 1.74E-4 | 0.01205 |
| 1.727E-4 | 0.01197 |
| 1.718E-4 | 0.01196 |
| 1.711E-4 | 0.01195 |
| 1.701E-4 | 0.01191 |
| 1.694E-4 | 0.01194 |
| 1.693E-4 | 0.01195 |
| 1.683E-4 | 0.01199 |
| 1.673E-4 | 0.01197 |
| 1.677E-4 | 0.01204 |
| 1.661E-4 | 0.01204 |
| 1.654E-4 | 0.01199 |
| 1.645E-4 | 0.01199 |
| 1.64E-4 | 0.01193 |
| 1.635E-4 | 0.01198 |
| 1.629E-4 | 0.01189 |
| 1.625E-4 | 0.01186 |
| 1.611E-4 | 0.01188 |
| 1.604E-4 | 0.01186 |
| 1.593E-4 | 0.0118 |
| 1.589E-4 | 0.01184 |
| 1.586E-4 | 0.01185 |
| 1.576E-4 | 0.01187 |
| 1.568E-4 | 0.01189 |
| 1.56E-4 | 0.01189 |
| 1.556E-4 | 0.01184 |
| 1.547E-4 | 0.01179 |
| 1.544E-4 | 0.01182 |
| 1.538E-4 | 0.01177 |
| 1.521E-4 | 0.01175 |
| 1.518E-4 | 0.01179 |
| 1.511E-4 | 0.01178 |
| 1.507E-4 | 0.01174 |
| 1.491E-4 | 0.01175 |
| 1.492E-4 | 0.01173 |
| 1.479E-4 | 0.01166 |
| 1.472E-4 | 0.01164 |
| 1.469E-4 | 0.01162 |
| 1.459E-4 | 0.01163 |
| 1.454E-4 | 0.0116 |
| 1.445E-4 | 0.01157 |
| 1.439E-4 | 0.01153 |
| 1.429E-4 | 0.01147 |
| 1.426E-4 | 0.01141 |
| 1.42E-4 | 0.01141 |
| 1.409E-4 | 0.01141 |
| 1.403E-4 | 0.01136 |
| 1.391E-4 | 0.01137 |
| 1.386E-4 | 0.01131 |
| 1.379E-4 | 0.01133 |
| 1.37E-4 | 0.01126 |
| 1.367E-4 | 0.01122 |
| 1.358E-4 | 0.01119 |
| 1.35E-4 | 0.01116 |
| 1.342E-4 | 0.01117 |
| 1.335E-4 | 0.01118 |
| 1.324E-4 | 0.01124 |
| 1.315E-4 | 0.01124 |
| 1.31E-4 | 0.01126 |
| 1.304E-4 | 0.01122 |
| 1.297E-4 | 0.01128 |
| 1.284E-4 | 0.01137 |
| 1.28E-4 | 0.01141 |
| 1.266E-4 | 0.01149 |
| 1.268E-4 | 0.01162 |
| 1.26E-4 | 0.01172 |
| 1.246E-4 | 0.0118 |
| 1.246E-4 | 0.01195 |
| 1.237E-4 | 0.012 |
| 1.233E-4 | 0.01214 |
| 1.222E-4 | 0.01226 |
| 1.217E-4 | 0.01244 |
| 1.215E-4 | 0.01256 |
| 1.201E-4 | 0.01268 |
| 1.199E-4 | 0.01267 |
| 1.19E-4 | 0.01268 |
| 1.183E-4 | 0.01285 |
| 1.176E-4 | 0.01288 |
| 1.17E-4 | 0.01292 |
| 1.165E-4 | 0.0131 |
| 1.16E-4 | 0.01317 |
| 1.154E-4 | 0.01316 |
| 1.141E-4 | 0.01315 |
| 1.14E-4 | 0.01318 |
| 1.131E-4 | 0.01323 |
| 1.125E-4 | 0.01333 |
| 1.12E-4 | 0.01335 |
| 1.113E-4 | 0.01339 |
| 1.105E-4 | 0.01332 |
| 1.102E-4 | 0.01328 |
| 1.094E-4 | 0.0133 |
| 1.085E-4 | 0.01331 |
| 1.082E-4 | 0.0133 |
| 1.067E-4 | 0.01335 |
| 1.067E-4 | 0.01341 |
| 1.061E-4 | 0.01338 |
| 1.052E-4 | 0.01339 |
| 1.052E-4 | 0.01334 |
| 1.04E-4 | 0.01334 |
| 1.033E-4 | 0.01329 |
| 1.026E-4 | 0.01326 |
| 1.022E-4 | 0.01322 |
| 1.018E-4 | 0.01324 |
| 1.008E-4 | 0.01325 |
| 1.003E-4 | 0.01322 |
| 9.96E-5 | 0.01323 |
| 9.91E-5 | 0.01309 |
| 9.79E-5 | 0.01307 |
| 9.75E-5 | 0.013 |
| 9.73E-5 | 0.013 |
| 9.6E-5 | 0.01295 |
| 9.6E-5 | 0.01292 |
| 9.52E-5 | 0.01291 |
| 9.44E-5 | 0.01291 |
| 9.37E-5 | 0.01279 |
| 9.31E-5 | 0.0127 |
| 9.25E-5 | 0.01266 |
| 9.19E-5 | 0.01262 |
| 9.12E-5 | 0.01255 |
| 9.01E-5 | 0.0125 |
| 8.99E-5 | 0.01233 |
| 8.89E-5 | 0.01221 |
| 8.78E-5 | 0.01217 |
| 8.8E-5 | 0.01208 |
| 8.68E-5 | 0.01196 |
| 8.62E-5 | 0.01191 |
| 8.54E-5 | 0.01182 |
| 8.49E-5 | 0.01173 |
| 8.33E-5 | 0.01158 |
| 8.33E-5 | 0.01139 |
| 8.25E-5 | 0.0113 |
| 8.18E-5 | 0.01126 |
| 8.13E-5 | 0.01112 |
| 7.98E-5 | 0.01103 |
| 7.9E-5 | 0.01093 |
| 7.85E-5 | 0.01077 |
| 7.77E-5 | 0.01062 |
| 7.75E-5 | 0.01057 |
| 7.64E-5 | 0.01048 |
| 7.53E-5 | 0.01027 |
| 7.45E-5 | 0.01016 |
| 7.35E-5 | 0.00997 |
| 7.23E-5 | 0.00973 |
| 7.18E-5 | 0.00949 |
| 7.01E-5 | 0.00925 |
| 6.94E-5 | 0.0089 |
| 6.9E-5 | 0.00848 |
| 6.69E-5 | 0.00794 |
| 6.58E-5 | 0.00737 |
| 6.43E-5 | 0.00677 |
| 6.29E-5 | 0.00613 |
| 6.07E-5 | 0.0057 |
| 6.05E-5 | 0.00539 |
| 5.91E-5 | 0.0051 |
| 5.65E-5 | 0.00488 |
| 5.5E-5 | 0.00466 |
| 5.23E-5 | 0.00453 |
| 5.06E-5 | 0.00454 |
| 4.96E-5 | 0.00465 |
| 4.8E-5 | 0.00501 |
| 4.69E-5 | 0.00533 |
| 4.57E-5 | 0.00578 |
| 4.5E-5 | 0.00652 |
| 4.34E-5 | 0.00727 |
| 4.3E-5 | 0.00841 |
| 4.21E-5 | 0.00991 |
| 4.14E-5 | 0.01178 |
| 4.14E-5 | 0.01359 |
| 4.04E-5 | 0.01565 |
| 4.02E-5 | 0.0176 |
| 3.97E-5 | 0.01954 |
| 3.93E-5 | 0.0212 |
| 3.88E-5 | 0.02195 |
| 3.87E-5 | 0.02232 |
| 3.79E-5 | 0.02309 |
| 3.76E-5 | 0.02309 |
| 3.73E-5 | 0.02363 |
| 3.65E-5 | 0.02382 |
| 3.66E-5 | 0.02346 |
| 3.61E-5 | 0.02377 |
| 3.57E-5 | 0.02422 |
| 3.56E-5 | 0.02438 |
| 3.5E-5 | 0.02495 |
| 3.5E-5 | 0.02505 |
| 3.43E-5 | 0.02496 |
| 3.41E-5 | 0.02533 |
| 3.41E-5 | 0.02535 |
| 3.34E-5 | 0.02531 |
| 3.3E-5 | 0.02547 |
| 3.25E-5 | 0.02545 |
| 3.23E-5 | 0.02525 |
| 3.18E-5 | 0.0255 |
| 3.19E-5 | 0.02517 |
| 3.13E-5 | 0.02545 |
| 3.1E-5 | 0.02573 |
| 3.03E-5 | 0.02563 |
| 3.02E-5 | 0.02592 |
| 3.01E-5 | 0.02597 |
| 2.95E-5 | 0.02576 |
| 2.99E-5 | 0.02602 |
| 2.9E-5 | 0.02614 |
| 2.86E-5 | 0.02622 |
| 2.79E-5 | 0.02626 |
| 2.8E-5 | 0.02623 |
| 2.83E-5 | 0.02611 |
| 2.7E-5 | 0.02618 |
| 2.72E-5 | 0.02649 |
| 2.68E-5 | 0.02613 |
| 2.64E-5 | 0.02617 |
| 2.62E-5 | 0.02604 |
| 2.57E-5 | 0.02583 |
| 2.55E-5 | 0.02601 |
| 2.48E-5 | 0.0262 |
| 2.5E-5 | 0.02613 |
| 2.42E-5 | 0.02632 |
| 2.41E-5 | 0.02631 |
| 2.42E-5 | 0.02616 |
| 2.34E-5 | 0.02636 |
| 2.3E-5 | 0.02629 |
| 2.29E-5 | 0.0262 |
| 2.24E-5 | 0.02633 |
| 2.18E-5 | 0.02641 |
| 2.17E-5 | 0.02631 |
| 2.12E-5 | 0.02613 |
| 2.12E-5 | 0.02623 |
| 2.11E-5 | 0.02628 |
| 2.06E-5 | 0.02633 |
| 2.07E-5 | 0.026 |
| 1.98E-5 | 0.02612 |
| 1.94E-5 | 0.02611 |
| 1.93E-5 | 0.02601 |
| 1.89E-5 | 0.02637 |
| 1.86E-5 | 0.02637 |
| 1.86E-5 | 0.02657 |
| 1.76E-5 | 0.02665 |
| 1.74E-5 | 0.02648 |
| 1.72E-5 | 0.02614 |
| 1.69E-5 | 0.02593 |
| 1.65E-5 | 0.02569 |
| 1.66E-5 | 0.0257 |
| 1.61E-5 | 0.026 |
| 1.61E-5 | 0.02592 |
| 1.54E-5 | 0.02595 |
| 1.47E-5 | 0.02607 |
| 1.45E-5 | 0.02612 |
| 1.46E-5 | 0.02579 |
| 1.37E-5 | 0.02599 |
| 1.39E-5 | 0.02602 |
| 1.36E-5 | 0.02599 |
| 1.28E-5 | 0.02575 |
| 1.28E-5 | 0.02556 |
| 1.23E-5 | 0.02557 |
| 1.18E-5 | 0.02543 |
| 1.16E-5 | 0.02512 |
| 1.14E-5 | 0.02513 |
| 1.1E-5 | 0.02492 |
| 1.05E-5 | 0.02485 |
| 1.02E-5 | 0.02513 |
| 9.7E-6 | 0.02482 |
| 9.8E-6 | 0.02475 |

```
9.1E-6  0.02463
9E-6    0.02443
8.6E-6  0.02396
7.7E-6  0.02394
7.9E-6  0.02381
7.3E-6  0.02364
7E-6    0.02315
6.7E-6  0.02266
6.2E-6  0.02228
6.1E-6  0.02215
5.4E-6  0.02188
5.4E-6  0.02185
4.5E-6  0.02159
4.7E-6  0.02164
4E-6    0.02176
3.6E-6  0.02223
3.8E-6  0.02263
2.6E-6  0.02353
2.7E-6  0.02482
2.5E-6  0.02639
2.3E-6  0.02868
1.6E-6  0.03086
1.6E-6  0.03345
1.3E-6  0.03604
1.2E-6  0.03895
1E-6    0.04326
7E-7    0.04708
7E-7    0.0495
1E-6    0.05039
7E-7    0.04864
8E-7    0.04959
6E-7    0.05138
3E-7    0.05129
4E-7    0.05055
0       0.04857
-2E-7   0.04542
-4E-7   0.04222
-8E-7   0.03816
-1.4E-6     0.03417
-1.6E-6     0.03136
-2E-6   0.02902
-2E-6   0.02686
-2.4E-6     0.02525
-2.9E-6     0.02388
-3E-6   0.02294
-3.4E-6     0.02222
```

| | |
|---|---|
| -3.8E-6 | 0.02172 |
| -3.7E-6 | 0.02151 |
| -4.5E-6 | 0.02161 |
| -5E-6 | 0.02166 |
| -5.3E-6 | 0.02163 |
| -5.9E-6 | 0.02194 |
| -6.2E-6 | 0.02218 |
| -6.6E-6 | 0.02237 |
| -6.8E-6 | 0.02281 |
| -7.3E-6 | 0.02329 |
| -7.4E-6 | 0.02341 |
| -7.8E-6 | 0.02373 |
| -8.1E-6 | 0.02403 |
| -8.6E-6 | 0.02421 |
| -9E-6 | 0.02427 |
| -9.4E-6 | 0.02418 |
| -9.7E-6 | 0.02418 |
| -1.01E-5 | 0.02434 |
| -1.03E-5 | 0.02465 |
| -1.04E-5 | 0.02492 |
| -1.05E-5 | 0.02532 |
| -1.15E-5 | 0.02519 |
| -1.17E-5 | 0.02515 |
| -1.21E-5 | 0.02536 |
| -1.28E-5 | 0.02565 |
| -1.24E-5 | 0.02589 |
| -1.32E-5 | 0.02611 |
| -1.31E-5 | 0.02601 |
| -1.37E-5 | 0.02613 |
| -1.38E-5 | 0.02612 |
| -1.44E-5 | 0.02597 |
| -1.47E-5 | 0.026 |
| -1.48E-5 | 0.02563 |
| -1.54E-5 | 0.02568 |
| -1.55E-5 | 0.02598 |
| -1.54E-5 | 0.02604 |
| -1.63E-5 | 0.02619 |
| -1.66E-5 | 0.02631 |
| -1.68E-5 | 0.02634 |
| -1.73E-5 | 0.02634 |
| -1.77E-5 | 0.02644 |
| -1.78E-5 | 0.0264 |
| -1.84E-5 | 0.02613 |
| -1.84E-5 | 0.02646 |
| -1.9E-5 | 0.02652 |
| -1.9E-5 | 0.02639 |
| -1.93E-5 | 0.02644 |
| -1.96E-5 | 0.02644 |
| -2.04E-5 | 0.02634 |
| -2.05E-5 | 0.02639 |
| -2.06E-5 | 0.02628 |
| -2.14E-5 | 0.02638 |
| -2.14E-5 | 0.02632 |
| -2.17E-5 | 0.02625 |
| -2.22E-5 | 0.02648 |
| -2.22E-5 | 0.02652 |
| -2.27E-5 | 0.02671 |
| -2.3E-5 | 0.02672 |
| -2.37E-5 | 0.02663 |
| -2.4E-5 | 0.02629 |
| -2.44E-5 | 0.02624 |
| -2.43E-5 | 0.02631 |
| -2.45E-5 | 0.02607 |
| -2.51E-5 | 0.02601 |
| -2.54E-5 | 0.02605 |
| -2.6E-5 | 0.02606 |
| -2.61E-5 | 0.0263 |
| -2.63E-5 | 0.02636 |
| -2.69E-5 | 0.02624 |
| -2.7E-5 | 0.0264 |
| -2.74E-5 | 0.02653 |
| -2.76E-5 | 0.0262 |
| -2.79E-5 | 0.02615 |
| -2.85E-5 | 0.02606 |
| -2.9E-5 | 0.02632 |
| -2.87E-5 | 0.02607 |
| -2.91E-5 | 0.02577 |
| -2.96E-5 | 0.02561 |
| -3E-5 | 0.02575 |
| -3.03E-5 | 0.02579 |
| -3.04E-5 | 0.02566 |
| -3.15E-5 | 0.02556 |
| -3.11E-5 | 0.02563 |
| -3.16E-5 | 0.02531 |
| -3.21E-5 | 0.02516 |
| -3.23E-5 | 0.02551 |
| -3.26E-5 | 0.02545 |
| -3.3E-5 | 0.02521 |
| -3.35E-5 | 0.02529 |
| -3.39E-5 | 0.02516 |
| -3.42E-5 | 0.02511 |
| -3.43E-5 | 0.02517 |
| -3.43E-5 | 0.02473 |
| -3.51E-5 | 0.0246 |
| -3.54E-5 | 0.02469 |
| -3.54E-5 | 0.02447 |
| -3.62E-5 | 0.0244 |
| -3.61E-5 | 0.02391 |
| -3.69E-5 | 0.02339 |
| -3.75E-5 | 0.02362 |
| -3.76E-5 | 0.02321 |
| -3.82E-5 | 0.02284 |
| -3.87E-5 | 0.02265 |
| -3.89E-5 | 0.02161 |
| -3.99E-5 | 0.02024 |
| -4.03E-5 | 0.01859 |
| -4.09E-5 | 0.01621 |
| -4.2E-5 | 0.01352 |
| -4.21E-5 | 0.01167 |
| -4.39E-5 | 0.01015 |
| -4.5E-5 | 0.00889 |
| -4.63E-5 | 0.00797 |
| -4.8E-5 | 0.00689 |
| -4.97E-5 | 0.0061 |
| -5.05E-5 | 0.00532 |
| -5.22E-5 | 0.00484 |
| -5.47E-5 | 0.00462 |
| -5.53E-5 | 0.00451 |
| -5.87E-5 | 0.00453 |
| -5.95E-5 | 0.00469 |
| -6.14E-5 | 0.00486 |
| -6.31E-5 | 0.00507 |
| -6.35E-5 | 0.00534 |
| -6.52E-5 | 0.00563 |
| -6.66E-5 | 0.00611 |
| -6.78E-5 | 0.00662 |
| -6.91E-5 | 0.00718 |
| -7.02E-5 | 0.00774 |
| -7.08E-5 | 0.00832 |
| -7.21E-5 | 0.00879 |
| -7.29E-5 | 0.00915 |
| -7.34E-5 | 0.00939 |
| -7.45E-5 | 0.0096 |
| -7.52E-5 | 0.00988 |
| -7.61E-5 | 0.01002 |
| -7.66E-5 | 0.0102 |
| -7.85E-5 | 0.01035 |
| -7.87E-5 | 0.01051 |
| -7.86E-5 | 0.01065 |
| -8E-5 | 0.01083 |
| -7.99E-5 | 0.01089 |
| -8.15E-5 | 0.01098 |
| -8.24E-5 | 0.0111 |
| -8.32E-5 | 0.01119 |
| -8.37E-5 | 0.0113 |
| -8.44E-5 | 0.01143 |
| -8.51E-5 | 0.01153 |
| -8.63E-5 | 0.01166 |
| -8.66E-5 | 0.01175 |
| -8.72E-5 | 0.0119 |
| -8.82E-5 | 0.01195 |
| -8.86E-5 | 0.01197 |
| -8.95E-5 | 0.01203 |
| -9.05E-5 | 0.01213 |
| -9.07E-5 | 0.01222 |
| -9.13E-5 | 0.01236 |
| -9.22E-5 | 0.01249 |
| -9.3E-5 | 0.01247 |
| -9.36E-5 | 0.01264 |
| -9.45E-5 | 0.01274 |
| -9.52E-5 | 0.01285 |
| -9.52E-5 | 0.01287 |
| -9.66E-5 | 0.01288 |
| -9.68E-5 | 0.01289 |
| -9.82E-5 | 0.01295 |
| -9.78E-5 | 0.01302 |
| -9.88E-5 | 0.01309 |
| -9.94E-5 | 0.01312 |
| -9.99E-5 | 0.01318 |
| -1.007E-4 | 0.01312 |
| -1.016E-4 | 0.0132 |
| -1.025E-4 | 0.01319 |
| -1.021E-4 | 0.01318 |
| -1.04E-4 | 0.01327 |
| -1.04E-4 | 0.01327 |
| -1.044E-4 | 0.01335 |
| -1.057E-4 | 0.01332 |
| -1.059E-4 | 0.01332 |
| -1.067E-4 | 0.01325 |
| -1.073E-4 | 0.01333 |
| -1.077E-4 | 0.01338 |
| -1.08E-4 | 0.01339 |
| -1.092E-4 | 0.01337 |
| -1.101E-4 | 0.01346 |
| -1.101E-4 | 0.01342 |
| -1.115E-4 | 0.01328 |
| -1.118E-4 | 0.01325 |
| -1.124E-4 | 0.01327 |
| -1.132E-4 | 0.0133 |
| -1.139E-4 | 0.0132 |
| -1.144E-4 | 0.01315 |
| -1.144E-4 | 0.01322 |
| -1.156E-4 | 0.01319 |
| -1.16E-4 | 0.01318 |
| -1.168E-4 | 0.01309 |
| -1.177E-4 | 0.01301 |
| -1.19E-4 | 0.01296 |
| -1.192E-4 | 0.01292 |
| -1.194E-4 | 0.01289 |
| -1.203E-4 | 0.01277 |
| -1.212E-4 | 0.01274 |
| -1.214E-4 | 0.01262 |
| -1.225E-4 | 0.01252 |
| -1.236E-4 | 0.01246 |
| -1.238E-4 | 0.01231 |
| -1.244E-4 | 0.01218 |
| -1.253E-4 | 0.0122 |
| -1.257E-4 | 0.012 |
| -1.267E-4 | 0.01187 |
| -1.276E-4 | 0.01179 |
| -1.282E-4 | 0.01168 |
| -1.293E-4 | 0.01154 |
| -1.293E-4 | 0.01143 |
| -1.303E-4 | 0.01132 |
| -1.309E-4 | 0.01128 |
| -1.314E-4 | 0.01121 |
| -1.324E-4 | 0.0111 |
| -1.337E-4 | 0.01106 |
| -1.34E-4 | 0.01108 |
| -1.352E-4 | 0.01109 |
| -1.359E-4 | 0.01111 |
| -1.36E-4 | 0.01117 |
| -1.372E-4 | 0.0112 |
| -1.375E-4 | 0.01125 |
| -1.386E-4 | 0.01127 |
| -1.391E-4 | 0.01128 |
| -1.403E-4 | 0.01131 |
| -1.405E-4 | 0.01138 |
| -1.414E-4 | 0.01138 |
| -1.424E-4 | 0.01142 |
| -1.427E-4 | 0.01148 |
| -1.441E-4 | 0.01154 |
| -1.446E-4 | 0.01157 |
| -1.455E-4 | 0.01152 |
| -1.464E-4 | 0.01152 |
| -1.466E-4 | 0.01157 |
| -1.472E-4 | 0.01159 |
| -1.483E-4 | 0.01165 |
| -1.491E-4 | 0.01164 |
| -1.496E-4 | 0.0116 |
| -1.509E-4 | 0.01169 |
| -1.511E-4 | 0.01168 |
| -1.513E-4 | 0.01169 |
| -1.525E-4 | 0.01163 |
| -1.532E-4 | 0.01164 |
| -1.541E-4 | 0.01169 |
| -1.546E-4 | 0.01173 |
| -1.559E-4 | 0.01177 |
| -1.558E-4 | 0.01178 |
| -1.572E-4 | 0.01179 |
| -1.576E-4 | 0.01184 |
| -1.578E-4 | 0.01185 |
| -1.594E-4 | 0.0119 |
| -1.598E-4 | 0.01191 |
| -1.604E-4 | 0.01188 |
| -1.609E-4 | 0.01184 |
| -1.621E-4 | 0.01189 |
| -1.626E-4 | 0.01195 |
| -1.629E-4 | 0.01192 |
| -1.636E-4 | 0.01195 |
| -1.644E-4 | 0.01198 |
| -1.658E-4 | 0.01199 |
| -1.654E-4 | 0.01194 |
| -1.671E-4 | 0.01196 |
| -1.675E-4 | 0.01199 |
| -1.678E-4 | 0.01202 |
| -1.693E-4 | 0.01196 |
| -1.695E-4 | 0.01203 |
| -1.703E-4 | 0.01202 |
| -1.711E-4 | 0.012 |
| -1.717E-4 | 0.01199 |
| -1.724E-4 | 0.01193 |
| -1.73E-4 | 0.012 |
| -1.739E-4 | 0.01199 |
| -1.745E-4 | 0.01197 |
| -1.751E-4 | 0.01198 |

| A | B |
|---|---|
| -1.758E-4 | 0.01209 |
| -1.765E-4 | 0.01207 |
| -1.777E-4 | 0.01204 |
| -1.781E-4 | 0.01203 |
| -1.786E-4 | 0.01203 |

dIdV124mK.dat

| A | B |
|---|---|
| 1.823E-4 | 0.01213 |
| 1.818E-4 | 0.01208 |
| 1.806E-4 | 0.01207 |
| 1.805E-4 | 0.01205 |
| 1.794E-4 | 0.01213 |
| 1.787E-4 | 0.01214 |
| 1.786E-4 | 0.01201 |
| 1.772E-4 | 0.01204 |
| 1.77E-4 | 0.01213 |
| 1.763E-4 | 0.01211 |
| 1.754E-4 | 0.01212 |
| 1.743E-4 | 0.01211 |
| 1.741E-4 | 0.01207 |
| 1.735E-4 | 0.01208 |
| 1.723E-4 | 0.01204 |
| 1.715E-4 | 0.01201 |
| 1.708E-4 | 0.01201 |
| 1.704E-4 | 0.01201 |
| 1.696E-4 | 0.012 |
| 1.689E-4 | 0.01202 |
| 1.685E-4 | 0.01199 |
| 1.674E-4 | 0.01194 |
| 1.67E-4 | 0.01205 |
| 1.659E-4 | 0.01199 |
| 1.652E-4 | 0.012 |
| 1.646E-4 | 0.01201 |
| 1.644E-4 | 0.01197 |
| 1.633E-4 | 0.01199 |
| 1.628E-4 | 0.01192 |
| 1.625E-4 | 0.01191 |
| 1.606E-4 | 0.01196 |
| 1.604E-4 | 0.01193 |
| 1.599E-4 | 0.01188 |
| 1.591E-4 | 0.01191 |
| 1.586E-4 | 0.01189 |
| 1.578E-4 | 0.01187 |
| 1.569E-4 | 0.01188 |
| 1.564E-4 | 0.01187 |
| 1.56E-4 | 0.01186 |
| 1.547E-4 | 0.01184 |
| 1.543E-4 | 0.01176 |
| 1.536E-4 | 0.01174 |
| 1.526E-4 | 0.01179 |
| 1.52E-4 | 0.01181 |
| 1.508E-4 | 0.0118 |
| 1.504E-4 | 0.01173 |
| 1.497E-4 | 0.01166 |
| 1.492E-4 | 0.01165 |
| 1.48E-4 | 0.01167 |
| 1.476E-4 | 0.01166 |
| 1.474E-4 | 0.01162 |
| 1.457E-4 | 0.01162 |
| 1.455E-4 | 0.01159 |
| 1.446E-4 | 0.01157 |
| 1.443E-4 | 0.01153 |
| 1.431E-4 | 0.01153 |
| 1.422E-4 | 0.01149 |
| 1.42E-4 | 0.0115 |
| 1.408E-4 | 0.0114 |
| 1.403E-4 | 0.01138 |
| 1.395E-4 | 0.0114 |
| 1.386E-4 | 0.0114 |
| 1.381E-4 | 0.01138 |
| 1.375E-4 | 0.01127 |
| 1.368E-4 | 0.0113 |
| 1.355E-4 | 0.01126 |
| 1.353E-4 | 0.01119 |
| 1.333E-4 | 0.01122 |
| 1.336E-4 | 0.01117 |
| 1.329E-4 | 0.01112 |
| 1.315E-4 | 0.0112 |
| 1.315E-4 | 0.0112 |
| 1.303E-4 | 0.01127 |
| 1.296E-4 | 0.01133 |
| 1.29E-4 | 0.01136 |
| 1.281E-4 | 0.01149 |
| 1.271E-4 | 0.01156 |
| 1.269E-4 | 0.01161 |
| 1.262E-4 | 0.01165 |
| 1.254E-4 | 0.0118 |
| 1.249E-4 | 0.0119 |
| 1.235E-4 | 0.01199 |
| 1.231E-4 | 0.01216 |
| 1.229E-4 | 0.01227 |
| 1.216E-4 | 0.01244 |
| 1.211E-4 | 0.01255 |
| 1.206E-4 | 0.01259 |
| 1.199E-4 | 0.0127 |
| 1.188E-4 | 0.01276 |
| 1.188E-4 | 0.01279 |
| 1.179E-4 | 0.01292 |
| 1.17E-4 | 0.01294 |
| 1.172E-4 | 0.01303 |
| 1.161E-4 | 0.01312 |
| 1.155E-4 | 0.01319 |
| 1.145E-4 | 0.01309 |
| 1.139E-4 | 0.01317 |
| 1.132E-4 | 0.01326 |
| 1.127E-4 | 0.01326 |
| 1.117E-4 | 0.01329 |
| 1.118E-4 | 0.0133 |
| 1.11E-4 | 0.01328 |
| 1.101E-4 | 0.01328 |
| 1.095E-4 | 0.01336 |
| 1.086E-4 | 0.01332 |
| 1.084E-4 | 0.01335 |
| 1.072E-4 | 0.01337 |
| 1.067E-4 | 0.01336 |
| 1.065E-4 | 0.01338 |
| 1.056E-4 | 0.01334 |
| 1.05E-4 | 0.01335 |
| 1.042E-4 | 0.01341 |
| 1.037E-4 | 0.0134 |
| 1.025E-4 | 0.01341 |
| 1.022E-4 | 0.01332 |
| 1.019E-4 | 0.01327 |
| 1.009E-4 | 0.01322 |
| 1.006E-4 | 0.01314 |
| 9.92E-5 | 0.01312 |
| 9.91E-5 | 0.01308 |
| 9.79E-5 | 0.01303 |
| 9.78E-5 | 0.01306 |
| 9.73E-5 | 0.01305 |
| 9.67E-5 | 0.01298 |
| 9.55E-5 | 0.01298 |
| 9.51E-5 | 0.01295 |
| 9.47E-5 | 0.01278 |
| 9.33E-5 | 0.01269 |
| 9.33E-5 | 0.0127 |
| 9.29E-5 | 0.01262 |
| 9.12E-5 | 0.01257 |
| 9.14E-5 | 0.01256 |
| 9.05E-5 | 0.01249 |
| 8.97E-5 | 0.01241 |
| 8.91E-5 | 0.01226 |
| 8.84E-5 | 0.01212 |
| 8.77E-5 | 0.01204 |
| 8.67E-5 | 0.01194 |
| 8.68E-5 | 0.01184 |
| 8.54E-5 | 0.01179 |
| 8.44E-5 | 0.01169 |
| 8.4E-5 | 0.01159 |
| 8.34E-5 | 0.01146 |
| 8.26E-5 | 0.01134 |
| 8.19E-5 | 0.01125 |
| 8.06E-5 | 0.01115 |
| 7.95E-5 | 0.01103 |
| 7.98E-5 | 0.01091 |
| 7.82E-5 | 0.01076 |
| 7.77E-5 | 0.01065 |
| 7.74E-5 | 0.01054 |
| 7.66E-5 | 0.01041 |
| 7.52E-5 | 0.01025 |
| 7.46E-5 | 0.01014 |
| 7.36E-5 | 0.00993 |
| 7.26E-5 | 0.00975 |
| 7.21E-5 | 0.00948 |
| 7.07E-5 | 0.00916 |
| 6.92E-5 | 0.00875 |
| 6.91E-5 | 0.00841 |
| 6.76E-5 | 0.008 |
| 6.57E-5 | 0.00745 |
| 6.44E-5 | 0.00683 |
| 6.35E-5 | 0.00622 |
| 6.11E-5 | 0.00582 |
| 6.09E-5 | 0.0054 |
| 5.89E-5 | 0.00512 |
| 5.68E-5 | 0.00495 |
| 5.63E-5 | 0.00476 |
| 5.25E-5 | 0.00465 |
| 5.19E-5 | 0.00461 |
| 4.99E-5 | 0.00473 |
| 4.85E-5 | 0.00505 |
| 4.77E-5 | 0.00542 |
| 4.64E-5 | 0.00589 |
| 4.49E-5 | 0.00654 |
| 4.41E-5 | 0.00717 |
| 4.27E-5 | 0.00836 |
| 4.26E-5 | 0.00964 |
| 4.18E-5 | 0.01141 |
| 4.13E-5 | 0.01328 |
| 4.06E-5 | 0.01512 |
| 4.04E-5 | 0.01715 |
| 3.95E-5 | 0.01917 |
| 3.91E-5 | 0.0204 |
| 3.91E-5 | 0.02173 |
| 3.86E-5 | 0.02281 |
| 3.82E-5 | 0.02281 |
| 3.76E-5 | 0.02307 |
| 3.73E-5 | 0.02368 |
| 3.69E-5 | 0.02344 |
| 3.67E-5 | 0.02367 |
| 3.63E-5 | 0.02426 |
| 3.59E-5 | 0.02388 |
| 3.57E-5 | 0.02419 |
| 3.53E-5 | 0.0246 |
| 3.5E-5 | 0.02464 |
| 3.45E-5 | 0.02512 |
| 3.39E-5 | 0.02535 |
| 3.38E-5 | 0.02523 |
| 3.34E-5 | 0.02546 |
| 3.35E-5 | 0.02525 |
| 3.24E-5 | 0.02515 |
| 3.25E-5 | 0.02544 |
| 3.25E-5 | 0.02548 |
| 3.17E-5 | 0.02579 |
| 3.11E-5 | 0.0259 |
| 3.13E-5 | 0.02571 |
| 3.05E-5 | 0.02574 |
| 2.99E-5 | 0.02566 |
| 3.06E-5 | 0.02571 |
| 2.98E-5 | 0.02577 |
| 2.92E-5 | 0.02564 |
| 2.94E-5 | 0.02571 |
| 2.84E-5 | 0.02614 |
| 2.85E-5 | 0.0263 |
| 2.83E-5 | 0.02612 |
| 2.77E-5 | 0.02605 |
| 2.73E-5 | 0.026 |
| 2.73E-5 | 0.02598 |
| 2.66E-5 | 0.02612 |
| 2.64E-5 | 0.02615 |
| 2.59E-5 | 0.02631 |
| 2.57E-5 | 0.02637 |
| 2.57E-5 | 0.02638 |
| 2.51E-5 | 0.02649 |
| 2.49E-5 | 0.02666 |
| 2.46E-5 | 0.02632 |
| 2.38E-5 | 0.02629 |
| 2.35E-5 | 0.02636 |
| 2.35E-5 | 0.02639 |
| 2.34E-5 | 0.02623 |
| 2.27E-5 | 0.02625 |
| 2.24E-5 | 0.02638 |
| 2.19E-5 | 0.02656 |
| 2.19E-5 | 0.02638 |
| 2.14E-5 | 0.02653 |
| 2.11E-5 | 0.02671 |
| 2.12E-5 | 0.02675 |
| 2.05E-5 | 0.02658 |
| 2.03E-5 | 0.02652 |
| 1.97E-5 | 0.02643 |
| 1.95E-5 | 0.02629 |
| 1.9E-5 | 0.02619 |
| 1.9E-5 | 0.02599 |
| 1.87E-5 | 0.026 |
| 1.84E-5 | 0.02623 |
| 1.78E-5 | 0.02627 |
| 1.78E-5 | 0.02651 |
| 1.71E-5 | 0.02664 |
| 1.69E-5 | 0.02651 |
| 1.63E-5 | 0.02651 |
| 1.67E-5 | 0.02665 |
| 1.61E-5 | 0.02642 |
| 1.59E-5 | 0.02639 |
| 1.53E-5 | 0.02623 |
| 1.47E-5 | 0.02626 |
| 1.44E-5 | 0.02629 |
| 1.48E-5 | 0.02587 |
| 1.38E-5 | 0.02574 |
| 1.39E-5 | 0.0258 |
| 1.35E-5 | 0.02588 |
| 1.29E-5 | 0.02579 |
| 1.27E-5 | 0.02543 |
| 1.23E-5 | 0.02536 |
| 1.2E-5 | 0.02554 |
| 1.14E-5 | 0.02561 |
| 1.15E-5 | 0.02537 |
| 1.11E-5 | 0.02534 |
| 1.05E-5 | 0.02526 |
| 1.01E-5 | 0.02506 |
| 1.02E-5 | 0.02487 |
| 9.5E-6 | 0.02474 |
| 9.2E-6 | 0.02471 |
| 8.8E-6 | 0.02464 |
| 8.6E-6 | 0.02444 |
| 8.3E-6 | 0.02399 |
| 8E-6 | 0.0236 |
| 7.3E-6 | 0.02357 |
| 7.2E-6 | 0.02344 |
| 6.4E-6 | 0.02338 |
| 6.6E-6 | 0.02318 |
| 6.7E-6 | 0.02263 |
| 5.2E-6 | 0.02233 |
| 5.2E-6 | 0.02225 |
| 4.1E-6 | 0.02221 |
| 4.4E-6 | 0.02239 |
| 3.9E-6 | 0.02225 |
| 4.2E-6 | 0.02254 |
| 3.7E-6 | 0.02314 |
| 3E-6 | 0.02375 |
| 3.3E-6 | 0.02455 |
| 2.2E-6 | 0.02584 |
| 2.3E-6 | 0.02753 |
| 2.2E-6 | 0.02967 |
| 1.9E-6 | 0.03253 |
| 1.6E-6 | 0.03501 |
| 1.4E-6 | 0.03776 |
| 1.2E-6 | 0.04103 |
| 9E-7 | 0.04325 |
| 4E-7 | 0.04475 |
| 1.3E-6 | 0.04613 |
| 7E-7 | 0.04582 |
| 8E-7 | 0.04529 |
| 6E-7 | 0.04593 |
| 2E-7 | 0.04703 |
| 5E-7 | 0.04616 |
| 0 | 0.04427 |
| 0 | 0.0417 |
| -3E-7 | 0.03819 |
| -8E-7 | 0.03531 |
| -8E-7 | 0.03217 |
| -1.3E-6 | 0.02987 |
| -1.9E-6 | 0.02807 |
| -2.2E-6 | 0.02653 |
| -2.3E-6 | 0.02527 |
| -2.8E-6 | 0.02447 |
| -3.2E-6 | 0.02341 |
| -3.7E-6 | 0.02281 |
| -3.8E-6 | 0.02232 |
| -4.1E-6 | 0.0222 |
| -4.6E-6 | 0.02224 |
| -4.7E-6 | 0.02215 |
| -5.2E-6 | 0.02219 |
| -5.7E-6 | 0.02249 |
| -6.3E-6 | 0.02281 |
| -6.5E-6 | 0.02306 |
| -6.9E-6 | 0.02357 |
| -7.1E-6 | 0.02378 |
| -7.6E-6 | 0.02396 |
| -7.8E-6 | 0.02428 |
| -8.7E-6 | 0.02444 |
| -8.6E-6 | 0.02471 |
| -9.1E-6 | 0.0246 |
| -8.9E-6 | 0.02465 |
| -9.5E-6 | 0.0247 |
| -1.02E-5 | 0.02488 |
| -9.9E-6 | 0.02483 |
| -1.05E-5 | 0.02487 |
| -1.12E-5 | 0.02516 |
| -1.13E-5 | 0.02536 |
| -1.13E-5 | 0.02543 |
| -1.23E-5 | 0.02518 |
| -1.21E-5 | 0.02559 |
| -1.25E-5 | 0.02593 |
| -1.31E-5 | 0.02591 |
| -1.3E-5 | 0.026 |
| -1.36E-5 | 0.02577 |
| -1.41E-5 | 0.02561 |
| -1.43E-5 | 0.02561 |
| -1.45E-5 | 0.02576 |
| -1.46E-5 | 0.0257 |
| -1.51E-5 | 0.02611 |
| -1.56E-5 | 0.0263 |
| -1.64E-5 | 0.02627 |
| -1.61E-5 | 0.02621 |
| -1.65E-5 | 0.02612 |
| -1.63E-5 | 0.02623 |
| -1.74E-5 | 0.02625 |
| -1.75E-5 | 0.02626 |
| -1.75E-5 | 0.02612 |
| -1.82E-5 | 0.02623 |
| -1.86E-5 | 0.02628 |
| -1.86E-5 | 0.02645 |
| -1.88E-5 | 0.02668 |
| -2E-5 | 0.02663 |
| -1.95E-5 | 0.02669 |
| -1.99E-5 | 0.02647 |
| -2.03E-5 | 0.02634 |
| -2.08E-5 | 0.0266 |
| -2.11E-5 | 0.02675 |
| -2.12E-5 | 0.02657 |
| -2.17E-5 | 0.02652 |
| -2.17E-5 | 0.0262 |
| -2.21E-5 | 0.0261 |
| -2.27E-5 | 0.02638 |
| -2.28E-5 | 0.02628 |
| -2.35E-5 | 0.02638 |
| -2.36E-5 | 0.02658 |
| -2.38E-5 | 0.02643 |
| -2.42E-5 | 0.02645 |
| -2.45E-5 | 0.02625 |
| -2.44E-5 | 0.02612 |
| -2.51E-5 | 0.02603 |
| -2.62E-5 | 0.02635 |
| -2.57E-5 | 0.02641 |
| -2.66E-5 | 0.02651 |
| -2.64E-5 | 0.02654 |
| -2.72E-5 | 0.02622 |
| -2.74E-5 | 0.02613 |
| -2.75E-5 | 0.02633 |
| -2.77E-5 | 0.02604 |
| -2.84E-5 | 0.02606 |
| -2.86E-5 | 0.02614 |
| -2.9E-5 | 0.02586 |
| -2.98E-5 | 0.02579 |
| -2.95E-5 | 0.02585 |
| -2.98E-5 | 0.02586 |
| -3.03E-5 | 0.02601 |
| -3.06E-5 | 0.026 |
| -3.1E-5 | 0.02582 |
| -3.14E-5 | 0.0255 |
| -3.17E-5 | 0.02555 |
| -3.18E-5 | 0.02569 |
| -3.24E-5 | 0.0256 |
| -3.27E-5 | 0.02548 |
| -3.27E-5 | 0.02541 |
| -3.33E-5 | 0.02525 |
| -3.36E-5 | 0.0251 |
| -3.42E-5 | 0.0249 |
| -3.41E-5 | 0.02472 |
| -3.48E-5 | 0.02468 |
| -3.52E-5 | 0.02453 |
| -3.53E-5 | 0.02415 |
| -3.59E-5 | 0.02443 |
| -3.63E-5 | 0.02435 |
| -3.65E-5 | 0.02401 |
| -3.72E-5 | 0.02388 |
| -3.75E-5 | 0.02308 |
| -3.79E-5 | 0.02273 |
| -3.84E-5 | 0.02242 |
| -3.88E-5 | 0.02134 |
| -3.93E-5 | 0.02053 |
| -3.99E-5 | 0.0191 |
| -4.06E-5 | 0.01677 |
| -4.14E-5 | 0.01502 |
| -4.24E-5 | 0.01236 |
| -4.3E-5 | 0.01074 |
| -4.4E-5 | 0.00973 |
| -4.48E-5 | 0.00858 |
| -4.72E-5 | 0.00772 |
| -4.73E-5 | 0.00676 |
| -5.07E-5 | 0.0059 |
| -5.16E-5 | 0.00522 |
| -5.23E-5 | 0.00483 |
| -5.55E-5 | 0.00461 |
| -5.64E-5 | 0.0046 |
| -5.84E-5 | 0.00467 |
| -5.99E-5 | 0.00489 |
| -6.17E-5 | 0.00513 |
| -6.25E-5 | 0.00538 |
| -6.4E-5 | 0.00564 |
| -6.53E-5 | 0.0059 |
| -6.64E-5 | 0.00629 |
| -6.81E-5 | 0.00669 |
| -6.85E-5 | 0.00726 |
| -7E-5 | 0.00778 |
| -7.09E-5 | 0.00823 |
| -7.19E-5 | 0.00858 |
| -7.28E-5 | 0.00904 |
| -7.36E-5 | 0.00934 |
| -7.41E-5 | 0.00957 |
| -7.5E-5 | 0.00985 |
| -7.6E-5 | 0.01008 |
| -7.66E-5 | 0.01028 |
| -7.78E-5 | 0.01038 |
| -7.84E-5 | 0.01054 |
| -7.87E-5 | 0.01063 |
| -8.02E-5 | 0.01076 |
| -8.06E-5 | 0.01087 |
| -8.17E-5 | 0.01093 |
| -8.21E-5 | 0.01104 |
| -8.32E-5 | 0.0112 |
| -8.38E-5 | 0.01131 |
| -8.43E-5 | 0.01136 |
| -8.48E-5 | 0.01146 |
| -8.59E-5 | 0.01161 |
| -8.69E-5 | 0.01175 |
| -8.75E-5 | 0.01187 |
| -8.8E-5 | 0.01196 |
| -8.88E-5 | 0.01212 |
| -8.89E-5 | 0.01222 |
| -9.02E-5 | 0.01225 |
| -9.06E-5 | 0.01236 |
| -9.11E-5 | 0.01244 |
| -9.16E-5 | 0.01249 |
| -9.37E-5 | 0.01256 |
| -9.38E-5 | 0.01266 |
| -9.45E-5 | 0.01271 |
| -9.46E-5 | 0.01276 |
| -9.52E-5 | 0.01278 |
| -9.59E-5 | 0.01282 |
| -9.67E-5 | 0.01291 |
| -9.79E-5 | 0.01296 |
| -9.82E-5 | 0.01295 |
| -9.9E-5 | 0.01297 |
| -9.93E-5 | 0.01306 |
| -9.99E-5 | 0.01319 |
| -1.008E-4 | 0.01323 |
| -1.011E-4 | 0.01321 |
| -1.025E-4 | 0.0132 |
| -1.03E-4 | 0.01322 |
| -1.032E-4 | 0.01324 |
| -1.04E-4 | 0.0132 |
| -1.043E-4 | 0.01327 |
| -1.049E-4 | 0.01331 |
| -1.059E-4 | 0.01332 |
| -1.065E-4 | 0.01335 |
| -1.071E-4 | 0.01333 |
| -1.08E-4 | 0.01332 |
| -1.087E-4 | 0.01329 |
| -1.088E-4 | 0.01335 |
| -1.096E-4 | 0.01338 |
| -1.104E-4 | 0.01339 |
| -1.111E-4 | 0.01339 |
| -1.111E-4 | 0.01337 |
| -1.134E-4 | 0.01339 |
| -1.126E-4 | 0.01331 |
| -1.139E-4 | 0.01326 |
| -1.145E-4 | 0.01318 |
| -1.147E-4 | 0.01318 |
| -1.154E-4 | 0.01314 |
| -1.159E-4 | 0.0131 |
| -1.172E-4 | 0.01302 |
| -1.171E-4 | 0.01303 |
| -1.185E-4 | 0.01296 |
| -1.192E-4 | 0.01283 |
| -1.19E-4 | 0.01283 |
| -1.204E-4 | 0.01279 |
| -1.207E-4 | 0.01264 |
| -1.218E-4 | 0.01259 |
| -1.222E-4 | 0.01246 |
| -1.234E-4 | 0.01236 |
| -1.234E-4 | 0.01225 |
| -1.242E-4 | 0.01218 |
| -1.248E-4 | 0.01213 |
| -1.257E-4 | 0.01189 |
| -1.273E-4 | 0.01176 |
| -1.271E-4 | 0.01176 |
| -1.283E-4 | 0.01164 |
| -1.287E-4 | 0.01157 |
| -1.29E-4 | 0.01145 |
| -1.306E-4 | 0.01131 |
| -1.309E-4 | 0.01129 |
| -1.312E-4 | 0.0112 |
| -1.324E-4 | 0.01116 |
| -1.333E-4 | 0.01113 |
| -1.338E-4 | 0.01118 |
| -1.351E-4 | 0.0112 |
| -1.356E-4 | 0.01124 |
| -1.362E-4 | 0.01121 |
| -1.367E-4 | 0.01127 |
| -1.378E-4 | 0.01127 |
| -1.385E-4 | 0.01126 |
| -1.394E-4 | 0.01125 |
| -1.402E-4 | 0.0113 |
| -1.406E-4 | 0.01129 |
| -1.412E-4 | 0.01135 |
| -1.425E-4 | 0.01143 |
| -1.428E-4 | 0.0115 |
| -1.441E-4 | 0.01158 |
| -1.443E-4 | 0.01156 |
| -1.454E-4 | 0.01163 |
| -1.462E-4 | 0.0116 |
| -1.459E-4 | 0.01156 |
| -1.475E-4 | 0.01159 |
| -1.479E-4 | 0.0116 |
| -1.487E-4 | 0.01162 |
| -1.494E-4 | 0.01168 |
| -1.509E-4 | 0.0117 |
| -1.508E-4 | 0.01167 |
| -1.514E-4 | 0.01168 |
| -1.526E-4 | 0.01175 |
| -1.53E-4 | 0.01172 |
| -1.536E-4 | 0.01168 |
| -1.544E-4 | 0.01173 |
| -1.553E-4 | 0.01178 |
| -1.558E-4 | 0.01181 |
| -1.57E-4 | 0.01183 |
| -1.574E-4 | 0.01185 |
| -1.578E-4 | 0.01188 |
| -1.59E-4 | 0.01189 |
| -1.592E-4 | 0.01188 |
| -1.605E-4 | 0.01184 |
| -1.61E-4 | 0.01183 |
| -1.622E-4 | 0.01183 |
| -1.622E-4 | 0.01182 |
| -1.628E-4 | 0.01185 |
| -1.638E-4 | 0.01187 |
| -1.644E-4 | 0.01192 |
| -1.653E-4 | 0.01185 |
| -1.657E-4 | 0.01195 |
| -1.668E-4 | 0.01195 |
| -1.675E-4 | 0.01197 |
| -1.678E-4 | 0.01197 |
| -1.688E-4 | 0.01197 |
| -1.695E-4 | 0.012 |
| -1.701E-4 | 0.01201 |
| -1.708E-4 | 0.01196 |
| -1.717E-4 | 0.012 |
| -1.72E-4 | 0.01205 |
| -1.726E-4 | 0.012 |
| -1.735E-4 | 0.01205 |
| -1.738E-4 | 0.01204 |
| -1.755E-4 | 0.01208 |
| -1.757E-4 | 0.01205 |
| -1.765E-4 | 0.01204 |
| -1.773E-4 | 0.01207 |
| -1.779E-4 | 0.01208 |
| -1.786E-4 | 0.01203 |
| -1.793E-4 | 0.01211 |

dIdV203mK.dat

| A | B |
|---|---|
| 1.83E-4 | 0.01206 |
| 1.822E-4 | 0.01208 |
| 1.81E-4 | 0.0121 |
| 1.801E-4 | 0.01208 |
| 1.793E-4 | 0.01213 |
| 1.791E-4 | 0.0121 |
| 1.782E-4 | 0.01206 |
| 1.775E-4 | 0.01204 |
| 1.776E-4 | 0.01204 |
| 1.757E-4 | 0.0121 |
| 1.759E-4 | 0.01209 |
| 1.743E-4 | 0.01202 |
| 1.735E-4 | 0.01202 |
| 1.737E-4 | 0.01205 |
| 1.726E-4 | 0.01206 |
| 1.72E-4 | 0.01204 |
| 1.713E-4 | 0.01204 |
| 1.705E-4 | 0.012 |
| 1.691E-4 | 0.01204 |
| 1.692E-4 | 0.01199 |
| 1.685E-4 | 0.012 |
| 1.675E-4 | 0.01198 |
| 1.673E-4 | 0.01189 |
| 1.661E-4 | 0.01195 |
| 1.656E-4 | 0.01199 |
| 1.646E-4 | 0.01195 |
| 1.644E-4 | 0.01193 |
| 1.635E-4 | 0.01193 |
| 1.626E-4 | 0.0119 |
| 1.621E-4 | 0.01192 |
| 1.614E-4 | 0.01192 |
| 1.612E-4 | 0.01188 |
| 1.597E-4 | 0.01186 |
| 1.592E-4 | 0.01187 |
| 1.588E-4 | 0.0119 |
| 1.575E-4 | 0.01189 |
| 1.573E-4 | 0.0119 |
| 1.56E-4 | 0.01189 |
| 1.56E-4 | 0.0118 |
| 1.55E-4 | 0.0118 |
| 1.54E-4 | 0.01184 |
| 1.538E-4 | 0.0118 |
| 1.524E-4 | 0.01179 |
| 1.519E-4 | 0.01179 |
| 1.514E-4 | 0.01176 |
| 1.51E-4 | 0.01168 |
| 1.498E-4 | 0.01174 |
| 1.494E-4 | 0.01167 |
| 1.481E-4 | 0.01166 |
| 1.476E-4 | 0.01158 |
| 1.474E-4 | 0.01162 |
| 1.46E-4 | 0.0116 |
| 1.465E-4 | 0.01162 |
| 1.444E-4 | 0.01156 |
| 1.439E-4 | 0.01157 |
| 1.428E-4 | 0.01157 |
| 1.43E-4 | 0.01154 |
| 1.417E-4 | 0.01154 |
| 1.412E-4 | 0.01152 |
| 1.409E-4 | 0.01144 |
| 1.389E-4 | 0.01142 |
| 1.384E-4 | 0.01132 |
| 1.385E-4 | 0.01123 |
| 1.374E-4 | 0.01122 |
| 1.366E-4 | 0.0112 |
| 1.354E-4 | 0.01117 |
| 1.348E-4 | 0.01118 |
| 1.344E-4 | 0.01114 |
| 1.333E-4 | 0.01118 |
| 1.322E-4 | 0.01121 |
| 1.322E-4 | 0.01119 |
| 1.312E-4 | 0.01123 |
| 1.303E-4 | 0.01125 |
| 1.3E-4 | 0.0113 |
| 1.287E-4 | 0.01139 |
| 1.281E-4 | 0.01143 |
| 1.269E-4 | 0.01148 |
| 1.266E-4 | 0.01156 |
| 1.263E-4 | 0.01161 |
| 1.249E-4 | 0.01172 |
| 1.246E-4 | 0.01186 |
| 1.243E-4 | 0.01198 |
| 1.231E-4 | 0.01213 |
| 1.225E-4 | 0.01226 |
| 1.218E-4 | 0.01229 |
| 1.217E-4 | 0.01239 |
| 1.205E-4 | 0.01247 |
| 1.198E-4 | 0.01254 |
| 1.189E-4 | 0.01266 |
| 1.187E-4 | 0.01279 |
| 1.176E-4 | 0.01283 |
| 1.174E-4 | 0.01293 |
| 1.168E-4 | 0.01301 |
| 1.158E-4 | 0.01304 |
| 1.151E-4 | 0.01309 |
| 1.141E-4 | 0.01314 |
| 1.142E-4 | 0.01319 |
| 1.128E-4 | 0.01327 |
| 1.123E-4 | 0.0133 |
| 1.12E-4 | 0.01327 |
| 1.11E-4 | 0.01328 |
| 1.107E-4 | 0.01329 |
| 1.098E-4 | 0.01324 |
| 1.094E-4 | 0.01328 |
| 1.085E-4 | 0.01334 |
| 1.077E-4 | 0.01336 |
| 1.076E-4 | 0.01338 |
| 1.068E-4 | 0.01342 |
| 1.063E-4 | 0.01338 |
| 1.053E-4 | 0.01344 |
| 1.053E-4 | 0.01341 |
| 1.04E-4 | 0.01338 |
| 1.039E-4 | 0.01338 |
| 1.032E-4 | 0.01343 |
| 1.021E-4 | 0.01335 |
| 1.02E-4 | 0.01333 |
| 1.009E-4 | 0.01328 |
| 1.001E-4 | 0.01326 |
| 9.95E-5 | 0.01323 |
| 9.93E-5 | 0.01321 |
| 9.83E-5 | 0.01321 |
| 9.8E-5 | 0.01312 |
| 9.72E-5 | 0.01312 |
| 9.63E-5 | 0.01309 |
| 9.64E-5 | 0.01304 |
| 9.51E-5 | 0.01298 |
| 9.48E-5 | 0.01292 |
| 9.39E-5 | 0.01283 |
| 9.32E-5 | 0.01271 |
| 9.25E-5 | 0.01267 |
| 9.22E-5 | 0.01262 |
| 9.15E-5 | 0.01252 |
| 9.01E-5 | 0.01253 |
| 9E-5 | 0.01237 |
| 8.89E-5 | 0.01228 |
| 8.81E-5 | 0.01222 |
| 8.81E-5 | 0.01212 |
| 8.64E-5 | 0.01201 |
| 8.64E-5 | 0.01197 |
| 8.52E-5 | 0.01184 |
| 8.46E-5 | 0.01162 |
| 8.38E-5 | 0.01152 |
| 8.32E-5 | 0.01143 |
| 8.24E-5 | 0.01131 |
| 8.18E-5 | 0.01125 |
| 8.06E-5 | 0.01117 |
| 7.99E-5 | 0.01107 |
| 7.91E-5 | 0.01095 |
| 7.79E-5 | 0.0108 |
| 7.75E-5 | 0.01069 |
| 7.78E-5 | 0.01062 |
| 7.64E-5 | 0.01051 |
| 7.48E-5 | 0.01034 |
| 7.39E-5 | 0.01022 |
| 7.33E-5 | 0.00997 |
| 7.21E-5 | 0.00982 |
| 7.15E-5 | 0.00959 |
| 7.07E-5 | 0.00928 |
| 6.91E-5 | 0.00892 |
| 6.84E-5 | 0.00861 |
| 6.78E-5 | 0.00824 |
| 6.6E-5 | 0.00765 |
| 6.42E-5 | 0.00709 |
| 6.35E-5 | 0.00653 |
| 6.12E-5 | 0.0061 |
| 6.07E-5 | 0.00569 |
| 5.95E-5 | 0.0054 |
| 5.68E-5 | 0.00518 |
| 5.55E-5 | 0.00502 |
| 5.39E-5 | 0.00487 |
| 5.18E-5 | 0.00481 |
| 5.04E-5 | 0.0048 |
| 4.94E-5 | 0.00497 |
| 4.73E-5 | 0.00524 |
| 4.66E-5 | 0.00556 |
| 4.5E-5 | 0.00626 |
| 4.34E-5 | 0.00695 |
| 4.33E-5 | 0.00783 |
| 4.23E-5 | 0.00882 |
| 4.09E-5 | 0.01007 |
| 4.17E-5 | 0.01129 |
| 4.04E-5 | 0.01304 |
| 3.98E-5 | 0.01518 |
| 3.91E-5 | 0.01743 |
| 3.89E-5 | 0.01945 |
| 3.88E-5 | 0.02052 |
| 3.83E-5 | 0.02174 |
| 3.79E-5 | 0.02231 |
| 3.75E-5 | 0.02238 |
| 3.71E-5 | 0.02288 |
| 3.63E-5 | 0.02371 |
| 3.64E-5 | 0.02381 |
| 3.57E-5 | 0.0242 |
| 3.5E-5 | 0.02444 |
| 3.56E-5 | 0.0244 |
| 3.5E-5 | 0.02478 |
| 3.43E-5 | 0.02486 |
| 3.41E-5 | 0.02477 |
| 3.38E-5 | 0.02506 |
| 3.35E-5 | 0.0249 |
| 3.31E-5 | 0.02492 |
| 3.29E-5 | 0.02533 |
| 3.23E-5 | 0.02542 |
| 3.2E-5 | 0.02513 |
| 3.14E-5 | 0.02555 |
| 3.13E-5 | 0.02547 |
| 3.12E-5 | 0.02525 |
| 3.07E-5 | 0.02554 |
| 3.04E-5 | 0.02546 |
| 2.99E-5 | 0.02579 |
| 2.97E-5 | 0.02576 |
| 2.95E-5 | 0.02564 |
| 2.92E-5 | 0.02582 |
| 2.86E-5 | 0.02595 |
| 2.85E-5 | 0.02584 |
| 2.79E-5 | 0.02626 |
| 2.77E-5 | 0.02603 |
| 2.76E-5 | 0.02601 |
| 2.69E-5 | 0.02598 |
| 2.68E-5 | 0.02615 |
| 2.62E-5 | 0.02602 |
| 2.58E-5 | 0.02606 |
| 2.56E-5 | 0.02628 |
| 2.56E-5 | 0.02634 |
| 2.54E-5 | 0.0266 |
| 2.49E-5 | 0.02692 |
| 2.51E-5 | 0.02638 |
| 2.41E-5 | 0.02618 |
| 2.38E-5 | 0.02619 |
| 2.35E-5 | 0.02612 |
| 2.29E-5 | 0.02648 |
| 2.29E-5 | 0.02659 |
| 2.24E-5 | 0.02667 |
| 2.25E-5 | 0.02652 |
| 2.17E-5 | 0.02628 |
| 2.18E-5 | 0.02631 |
| 2.12E-5 | 0.0263 |
| 2.11E-5 | 0.02628 |
| 2.08E-5 | 0.02631 |
| 2.04E-5 | 0.02643 |
| 1.99E-5 | 0.02655 |
| 1.97E-5 | 0.02644 |
| 1.95E-5 | 0.02663 |
| 1.9E-5 | 0.0265 |
| 1.89E-5 | 0.02646 |
| 1.86E-5 | 0.0266 |
| 1.79E-5 | 0.02649 |
| 1.79E-5 | 0.02651 |
| 1.7E-5 | 0.02641 |
| 1.73E-5 | 0.02603 |
| 1.67E-5 | 0.02608 |
| 1.64E-5 | 0.0263 |
| 1.58E-5 | 0.02622 |
| 1.6E-5 | 0.02597 |
| 1.55E-5 | 0.02603 |
| 1.52E-5 | 0.02604 |
| 1.5E-5 | 0.026 |
| 1.44E-5 | 0.02612 |
| 1.42E-5 | 0.02587 |
| 1.39E-5 | 0.02582 |
| 1.38E-5 | 0.02566 |
| 1.35E-5 | 0.02565 |
| 1.27E-5 | 0.02541 |
| 1.26E-5 | 0.02558 |
| 1.23E-5 | 0.02565 |
| 1.19E-5 | 0.02553 |
| 1.19E-5 | 0.02552 |
| 1.13E-5 | 0.02563 |
| 1.12E-5 | 0.02555 |
| 1.06E-5 | 0.02559 |
| 1.03E-5 | 0.02529 |
| 9.9E-6 | 0.02542 |
| 9.8E-6 | 0.02538 |
| 9E-6 | 0.02519 |
| 8.4E-6 | 0.02509 |
| 9E-6 | 0.02503 |
| 8.1E-6 | 0.02508 |
| 7.5E-6 | 0.02469 |
| 7.7E-6 | 0.0244 |
| 7.3E-6 | 0.02397 |
| 6.5E-6 | 0.02399 |
| 6.1E-6 | 0.02364 |
| 6.1E-6 | 0.02377 |
| 5.7E-6 | 0.02375 |
| 5.5E-6 | 0.02364 |
| 5.1E-6 | 0.02361 |
| 4.4E-6 | 0.02348 |
| 4.3E-6 | 0.0236 |
| 4.1E-6 | 0.024 |
| 3.6E-6 | 0.02426 |
| 3.2E-6 | 0.02445 |
| 3.2E-6 | 0.02514 |
| 2.6E-6 | 0.02616 |
| 2.4E-6 | 0.02761 |
| 1.7E-6 | 0.02906 |
| 1.8E-6 | 0.03023 |
| 1.9E-6 | 0.03126 |
| 1.2E-6 | 0.03257 |
| 1.2E-6 | 0.03437 |
| 5E-7 | 0.03544 |
| 5E-7 | 0.03647 |
| 1.3E-6 | 0.03706 |
| 1.1E-6 | 0.03704 |
| 3E-7 | 0.0372 |
| 6E-7 | 0.03683 |
| 1E-7 | 0.03694 |
| -1E-7 | 0.03692 |
| 2E-7 | 0.03618 |
| -2E-7 | 0.03517 |
| -6E-7 | 0.0339 |
| -1.5E-6 | 0.03221 |
| -1.3E-6 | 0.03003 |
| -1.8E-6 | 0.02866 |
| -1.9E-6 | 0.02711 |
| -2E-6 | 0.026 |
| -2.5E-6 | 0.02538 |
| -2.6E-6 | 0.02471 |
| -3.4E-6 | 0.02422 |
| -4E-6 | 0.02416 |
| -3.6E-6 | 0.0239 |
| -4.1E-6 | 0.02364 |
| -4.7E-6 | 0.02348 |
| -4.9E-6 | 0.02327 |
| -5.2E-6 | 0.02356 |
| -6.2E-6 | 0.02373 |
| -5.9E-6 | 0.02404 |
| -6.5E-6 | 0.0242 |
| -7.1E-6 | 0.02451 |
| -7.2E-6 | 0.0244 |
| -7.7E-6 | 0.02451 |
| -7.4E-6 | 0.02459 |
| -8.2E-6 | 0.02461 |
| -8.8E-6 | 0.02473 |
| -9E-6 | 0.02499 |
| -9.1E-6 | 0.02508 |
| -9.6E-6 | 0.02511 |
| -1E-5 | 0.02535 |
| -1.02E-5 | 0.02528 |
| -1.05E-5 | 0.0255 |
| -1.09E-5 | 0.02573 |
| -1.11E-5 | 0.02572 |
| -1.15E-5 | 0.02577 |
| -1.18E-5 | 0.02557 |
| -1.25E-5 | 0.02575 |
| -1.25E-5 | 0.0258 |
| -1.3E-5 | 0.02578 |
| -1.32E-5 | 0.02582 |
| -1.38E-5 | 0.02601 |
| -1.39E-5 | 0.02599 |
| -1.45E-5 | 0.02616 |
| -1.47E-5 | 0.02606 |
| -1.46E-5 | 0.02603 |
| -1.54E-5 | 0.02642 |
| -1.55E-5 | 0.02633 |
| -1.59E-5 | 0.02646 |
| -1.6E-5 | 0.02642 |
| -1.59E-5 | 0.0265 |
| -1.67E-5 | 0.02661 |
| -1.73E-5 | 0.02623 |
| -1.74E-5 | 0.02614 |
| -1.76E-5 | 0.02631 |
| -1.84E-5 | 0.02626 |
| -1.89E-5 | 0.02626 |
| -1.85E-5 | 0.02628 |
| -1.91E-5 | 0.02625 |
| -1.99E-5 | 0.0262 |
| -2.01E-5 | 0.02646 |
| -2.02E-5 | 0.02682 |
| -2.04E-5 | 0.02676 |
| -2.1E-5 | 0.02688 |
| -2.09E-5 | 0.02661 |
| -2.16E-5 | 0.02648 |
| -2.18E-5 | 0.02664 |
| -2.21E-5 | 0.02642 |
| -2.2E-5 | 0.02642 |
| -2.28E-5 | 0.02645 |
| -2.32E-5 | 0.02636 |
| -2.36E-5 | 0.02635 |
| -2.38E-5 | 0.02626 |
| -2.41E-5 | 0.02614 |
| -2.43E-5 | 0.02616 |
| -2.48E-5 | 0.02618 |
| -2.5E-5 | 0.02631 |
| -2.53E-5 | 0.02649 |
| -2.57E-5 | 0.02635 |
| -2.57E-5 | 0.02623 |
| -2.65E-5 | 0.02605 |
| -2.66E-5 | 0.02594 |
| -2.71E-5 | 0.02595 |
| -2.75E-5 | 0.02598 |
| -2.74E-5 | 0.026 |
| -2.83E-5 | 0.02592 |
| -2.82E-5 | 0.02589 |
| -2.89E-5 | 0.02595 |
| -2.87E-5 | 0.02602 |
| -2.93E-5 | 0.0257 |
| -2.98E-5 | 0.02571 |
| -2.99E-5 | 0.02585 |
| -3.07E-5 | 0.02575 |
| -3.02E-5 | 0.02589 |
| -3.15E-5 | 0.02589 |
| -3.16E-5 | 0.02589 |
| -3.22E-5 | 0.02564 |
| -3.21E-5 | 0.02518 |
| -3.25E-5 | 0.02487 |
| -3.28E-5 | 0.02515 |
| -3.31E-5 | 0.02509 |
| -3.34E-5 | 0.02503 |
| -3.39E-5 | 0.02511 |
| -3.43E-5 | 0.02477 |
| -3.45E-5 | 0.0249 |
| -3.45E-5 | 0.02471 |
| -3.55E-5 | 0.02454 |
| -3.54E-5 | 0.02462 |
| -3.6E-5 | 0.02433 |
| -3.63E-5 | 0.0239 |
| -3.7E-5 | 0.02367 |
| -3.73E-5 | 0.02325 |
| -3.73E-5 | 0.02277 |
| -3.77E-5 | 0.02257 |
| -3.83E-5 | 0.02188 |
| -3.91E-5 | 0.02125 |
| -3.95E-5 | 0.02025 |
| -4E-5 | 0.01868 |
| -4.1E-5 | 0.01633 |
| -4.16E-5 | 0.01415 |
| -4.28E-5 | 0.01171 |
| -4.33E-5 | 0.01003 |
| -4.43E-5 | 0.00892 |
| -4.57E-5 | 0.00786 |
| -4.74E-5 | 0.00713 |
| -4.78E-5 | 0.00638 |
| -5.11E-5 | 0.00575 |
| -5.21E-5 | 0.00525 |
| -5.29E-5 | 0.0049 |
| -5.57E-5 | 0.00475 |
| -5.72E-5 | 0.00477 |
| -5.88E-5 | 0.00485 |
| -6.03E-5 | 0.00497 |
| -6.24E-5 | 0.00513 |
| -6.26E-5 | 0.00539 |
| -6.41E-5 | 0.00565 |
| -6.6E-5 | 0.00599 |
| -6.64E-5 | 0.00641 |
| -6.8E-5 | 0.00694 |
| -6.9E-5 | 0.0075 |
| -7E-5 | 0.00803 |
| -7.11E-5 | 0.0085 |
| -7.19E-5 | 0.00893 |
| -7.3E-5 | 0.00927 |
| -7.32E-5 | 0.00951 |
| -7.47E-5 | 0.00974 |
| -7.48E-5 | 0.00992 |
| -7.65E-5 | 0.01012 |
| -7.7E-5 | 0.01027 |
| -7.78E-5 | 0.01043 |
| -7.85E-5 | 0.01055 |
| -7.93E-5 | 0.01075 |
| -7.97E-5 | 0.01085 |
| -8.05E-5 | 0.01097 |
| -8.22E-5 | 0.01106 |
| -8.2E-5 | 0.01118 |
| -8.36E-5 | 0.0112 |
| -8.4E-5 | 0.01137 |
| -8.44E-5 | 0.01144 |
| -8.52E-5 | 0.01153 |
| -8.6E-5 | 0.01169 |
| -8.67E-5 | 0.01184 |
| -8.71E-5 | 0.01192 |
| -8.84E-5 | 0.01202 |
| -8.88E-5 | 0.01212 |
| -8.89E-5 | 0.01219 |
| -9.04E-5 | 0.01225 |
| -9.07E-5 | 0.01239 |
| -9.16E-5 | 0.01242 |
| -9.23E-5 | 0.01256 |
| -9.35E-5 | 0.01263 |
| -9.33E-5 | 0.0126 |
| -9.4E-5 | 0.01273 |
| -9.48E-5 | 0.01283 |
| -9.54E-5 | 0.01284 |
| -9.61E-5 | 0.01283 |
| -9.67E-5 | 0.01285 |
| -9.77E-5 | 0.01296 |
| -9.82E-5 | 0.01305 |
| -9.94E-5 | 0.01304 |
| -9.96E-5 | 0.01307 |
| -1.003E-4 | 0.01309 |
| -1.009E-4 | 0.01301 |
| -1.011E-4 | 0.01319 |
| -1.022E-4 | 0.01336 |
| -1.024E-4 | 0.01327 |
| -1.038E-4 | 0.01328 |
| -1.038E-4 | 0.01324 |
| -1.045E-4 | 0.0132 |
| -1.056E-4 | 0.01318 |
| -1.057E-4 | 0.0132 |
| -1.067E-4 | 0.0133 |
| -1.073E-4 | 0.0133 |
| -1.078E-4 | 0.01327 |
| -1.085E-4 | 0.01329 |
| -1.091E-4 | 0.01336 |
| -1.095E-4 | 0.01332 |
| -1.103E-4 | 0.01339 |
| -1.111E-4 | 0.01339 |
| -1.112E-4 | 0.01342 |
| -1.125E-4 | 0.01328 |
| -1.128E-4 | 0.01327 |
| -1.139E-4 | 0.01322 |
| -1.145E-4 | 0.01311 |
| -1.152E-4 | 0.01318 |
| -1.157E-4 | 0.01321 |
| -1.162E-4 | 0.01308 |
| -1.168E-4 | 0.01305 |
| -1.176E-4 | 0.01298 |
| -1.184E-4 | 0.01296 |
| -1.187E-4 | 0.01292 |
| -1.193E-4 | 0.01279 |
| -1.201E-4 | 0.01282 |
| -1.209E-4 | 0.0127 |
| -1.218E-4 | 0.01258 |
| -1.223E-4 | 0.01242 |
| -1.229E-4 | 0.01235 |
| -1.24E-4 | 0.01231 |
| -1.238E-4 | 0.01219 |
| -1.253E-4 | 0.01204 |
| -1.257E-4 | 0.01188 |
| -1.268E-4 | 0.01179 |
| -1.272E-4 | 0.01172 |
| -1.286E-4 | 0.0116 |
| -1.287E-4 | 0.01151 |
| -1.292E-4 | 0.0114 |
| -1.3E-4 | 0.01139 |
| -1.309E-4 | 0.01132 |
| -1.319E-4 | 0.01127 |
| -1.324E-4 | 0.0113 |
| -1.336E-4 | 0.01132 |
| -1.336E-4 | 0.01121 |
| -1.355E-4 | 0.01121 |
| -1.35E-4 | 0.0112 |
| -1.359E-4 | 0.01119 |
| -1.376E-4 | 0.01123 |
| -1.38E-4 | 0.01121 |
| -1.384E-4 | 0.01125 |
| -1.393E-4 | 0.01131 |
| -1.402E-4 | 0.01133 |
| -1.404E-4 | 0.01134 |
| -1.414E-4 | 0.01138 |
| -1.424E-4 | 0.01146 |
| -1.426E-4 | 0.01147 |
| -1.443E-4 | 0.01154 |
| -1.445E-4 | 0.01156 |
| -1.458E-4 | 0.01152 |
| -1.462E-4 | 0.01164 |
| -1.465E-4 | 0.01165 |
| -1.472E-4 | 0.01159 |
| -1.48E-4 | 0.01164 |
| -1.495E-4 | 0.01171 |
| -1.493E-4 | 0.01173 |
| -1.508E-4 | 0.01177 |
| -1.511E-4 | 0.01177 |
| -1.514E-4 | 0.0117 |
| -1.526E-4 | 0.01176 |
| -1.532E-4 | 0.01174 |
| -1.541E-4 | 0.01183 |
| -1.545E-4 | 0.01181 |
| -1.554E-4 | 0.01178 |
| -1.559E-4 | 0.01179 |
| -1.569E-4 | 0.01181 |
| -1.578E-4 | 0.01185 |
| -1.579E-4 | 0.01186 |
| -1.59E-4 | 0.01184 |
| -1.596E-4 | 0.01184 |
| -1.609E-4 | 0.01192 |
| -1.607E-4 | 0.0119 |
| -1.625E-4 | 0.01198 |
| -1.629E-4 | 0.01192 |
| -1.631E-4 | 0.01198 |
| -1.64E-4 | 0.012 |
| -1.645E-4 | 0.01201 |
| -1.653E-4 | 0.012 |
| -1.66E-4 | 0.01199 |
| -1.672E-4 | 0.01195 |
| -1.673E-4 | 0.01197 |
| -1.679E-4 | 0.01199 |
| -1.69E-4 | 0.012 |
| -1.691E-4 | 0.01204 |
| -1.705E-4 | 0.01206 |
| -1.708E-4 | 0.01207 |
| -1.718E-4 | 0.01202 |
| -1.722E-4 | 0.01208 |
| -1.726E-4 | 0.01209 |
| -1.739E-4 | 0.01206 |
| -1.745E-4 | 0.01216 |
| -1.75E-4 | 0.01215 |
| -1.757E-4 | 0.01207 |
| -1.767E-4 | 0.01213 |
| -1.771E-4 | 0.01211 |
| -1.783E-4 | 0.01208 |
| -1.788E-4 | 0.01207 |
| -1.791E-4 | 0.01205 |

dIdV350mK.dat

| A | B |
|---|---|
| 1.822E-4 | 0.01219 |
| 1.822E-4 | 0.01219 |
| 1.816E-4 | 0.0122 |
| 1.811E-4 | 0.01223 |
| 1.798E-4 | 0.01221 |
| 1.792E-4 | 0.01227 |
| 1.792E-4 | 0.01222 |
| 1.782E-4 | 0.01216 |
| 1.772E-4 | 0.01214 |
| 1.769E-4 | 0.01222 |
| 1.762E-4 | 0.01217 |
| 1.749E-4 | 0.01219 |
| 1.747E-4 | 0.01223 |
| 1.745E-4 | 0.0122 |
| 1.733E-4 | 0.01216 |
| 1.727E-4 | 0.01221 |
| 1.719E-4 | 0.0122 |
| 1.712E-4 | 0.0122 |
| 1.702E-4 | 0.0122 |
| 1.696E-4 | 0.01226 |
| 1.694E-4 | 0.01221 |
| 1.686E-4 | 0.01224 |
| 1.676E-4 | 0.01223 |
| 1.669E-4 | 0.01223 |
| 1.663E-4 | 0.01217 |
| 1.651E-4 | 0.01215 |
| 1.644E-4 | 0.01215 |
| 1.645E-4 | 0.01214 |
| 1.63E-4 | 0.01213 |
| 1.628E-4 | 0.01215 |
| 1.623E-4 | 0.01215 |
| 1.612E-4 | 0.01213 |
| 1.602E-4 | 0.01211 |
| 1.599E-4 | 0.01211 |
| 1.593E-4 | 0.01212 |
| 1.583E-4 | 0.01212 |
| 1.58E-4 | 0.0121 |
| 1.565E-4 | 0.01208 |
| 1.568E-4 | 0.01207 |
| 1.554E-4 | 0.01204 |
| 1.553E-4 | 0.01207 |
| 1.547E-4 | 0.0121 |
| 1.543E-4 | 0.01211 |
| 1.531E-4 | 0.01209 |
| 1.524E-4 | 0.01204 |
| 1.519E-4 | 0.01204 |
| 1.506E-4 | 0.01206 |
| 1.502E-4 | 0.01191 |
| 1.493E-4 | 0.01193 |
| 1.486E-4 | 0.01197 |
| 1.482E-4 | 0.01192 |
| 1.473E-4 | 0.0119 |
| 1.468E-4 | 0.01189 |
| 1.457E-4 | 0.0119 |
| 1.449E-4 | 0.01189 |
| 1.442E-4 | 0.01189 |
| 1.438E-4 | 0.01189 |
| 1.429E-4 | 0.01188 |
| 1.422E-4 | 0.01192 |
| 1.423E-4 | 0.01188 |
| 1.406E-4 | 0.01187 |
| 1.403E-4 | 0.01184 |
| 1.392E-4 | 0.01173 |
| 1.39E-4 | 0.01175 |
| 1.381E-4 | 0.01174 |
| 1.374E-4 | 0.01174 |
| 1.364E-4 | 0.0117 |
| 1.356E-4 | 0.01166 |
| 1.355E-4 | 0.01166 |
| 1.342E-4 | 0.01167 |
| 1.338E-4 | 0.01166 |
| 1.331E-4 | 0.01153 |
| 1.322E-4 | 0.01147 |
| 1.321E-4 | 0.01143 |
| 1.301E-4 | 0.01142 |
| 1.297E-4 | 0.0114 |
| 1.292E-4 | 0.01139 |
| 1.282E-4 | 0.01141 |
| 1.278E-4 | 0.01148 |
| 1.27E-4 | 0.01147 |
| 1.264E-4 | 0.01145 |
| 1.25E-4 | 0.01141 |
| 1.25E-4 | 0.01139 |
| 1.236E-4 | 0.01141 |
| 1.232E-4 | 0.01147 |
| 1.228E-4 | 0.01151 |
| 1.221E-4 | 0.01168 |
| 1.21E-4 | 0.01186 |
| 1.205E-4 | 0.01192 |
| 1.195E-4 | 0.01197 |
| 1.189E-4 | 0.0121 |
| 1.181E-4 | 0.01223 |
| 1.179E-4 | 0.01236 |
| 1.167E-4 | 0.01247 |
| 1.164E-4 | 0.01252 |
| 1.157E-4 | 0.01262 |
| 1.152E-4 | 0.01275 |
| 1.141E-4 | 0.01286 |
| 1.138E-4 | 0.01295 |
| 1.125E-4 | 0.01303 |
| 1.122E-4 | 0.01312 |
| 1.12E-4 | 0.01319 |
| 1.111E-4 | 0.01315 |
| 1.104E-4 | 0.01317 |
| 1.096E-4 | 0.01318 |
| 1.091E-4 | 0.0133 |
| 1.083E-4 | 0.01335 |
| 1.081E-4 | 0.01326 |
| 1.074E-4 | 0.01331 |
| 1.064E-4 | 0.01342 |
| 1.061E-4 | 0.0135 |
| 1.051E-4 | 0.01343 |
| 1.044E-4 | 0.01349 |
| 1.038E-4 | 0.0134 |
| 1.03E-4 | 0.01335 |
| 1.026E-4 | 0.01332 |
| 1.019E-4 | 0.01333 |
| 1.016E-4 | 0.01335 |
| 1.004E-4 | 0.01339 |
| 1.007E-4 | 0.01335 |
| 9.96E-5 | 0.01332 |
| 9.87E-5 | 0.01335 |
| 9.84E-5 | 0.01334 |
| 9.73E-5 | 0.01318 |
| 9.72E-5 | 0.01323 |
| 9.58E-5 | 0.01324 |
| 9.57E-5 | 0.01325 |
| 9.48E-5 | 0.0133 |
| 9.44E-5 | 0.01331 |
| 9.37E-5 | 0.01317 |
| 9.29E-5 | 0.01311 |
| 9.26E-5 | 0.01307 |
| 9.16E-5 | 0.01309 |
| 9.1E-5 | 0.01311 |
| 9.03E-5 | 0.01309 |
| 8.97E-5 | 0.01298 |
| 8.92E-5 | 0.01301 |
| 8.88E-5 | 0.01291 |
| 8.75E-5 | 0.01275 |
| 8.67E-5 | 0.0127 |
| 8.68E-5 | 0.01272 |
| 8.54E-5 | 0.01263 |
| 8.48E-5 | 0.01255 |
| 8.46E-5 | 0.01246 |
| 8.36E-5 | 0.01238 |
| 8.32E-5 | 0.01237 |
| 8.24E-5 | 0.01229 |
| 8.16E-5 | 0.01224 |
| 8.05E-5 | 0.01225 |
| 8.03E-5 | 0.01209 |
| 7.95E-5 | 0.01203 |
| 7.91E-5 | 0.01196 |
| 7.83E-5 | 0.01188 |
| 7.7E-5 | 0.01179 |
| 7.67E-5 | 0.01166 |
| 7.58E-5 | 0.01165 |
| 7.49E-5 | 0.01163 |
| 7.44E-5 | 0.01158 |
| 7.39E-5 | 0.01147 |
| 7.3E-5 | 0.01149 |
| 7.22E-5 | 0.01148 |
| 7.16E-5 | 0.01138 |
| 7.01E-5 | 0.01135 |
| 7.03E-5 | 0.01126 |
| 6.91E-5 | 0.01118 |
| 6.83E-5 | 0.01105 |
| 6.79E-5 | 0.01094 |
| 6.66E-5 | 0.01085 |
| 6.58E-5 | 0.01071 |
| 6.49E-5 | 0.01054 |
| 6.43E-5 | 0.01035 |
| 6.29E-5 | 0.01016 |
| 6.27E-5 | 0.00991 |
| 6.15E-5 | 0.00972 |
| 6.07E-5 | 0.00953 |
| 6E-5 | 0.00929 |
| 5.84E-5 | 0.00905 |
| 5.78E-5 | 0.00881 |
| 5.67E-5 | 0.00854 |
| 5.54E-5 | 0.0083 |
| 5.55E-5 | 0.00809 |
| 5.39E-5 | 0.00789 |
| 5.26E-5 | 0.00767 |
| 5.14E-5 | 0.00752 |
| 5.04E-5 | 0.00734 |
| 4.84E-5 | 0.00718 |
| 4.81E-5 | 0.00708 |
| 4.76E-5 | 0.00707 |
| 4.53E-5 | 0.0071 |
| 4.46E-5 | 0.0071 |
| 4.32E-5 | 0.00717 |
| 4.24E-5 | 0.00746 |
| 4.08E-5 | 0.00774 |
| 4.05E-5 | 0.00808 |
| 3.94E-5 | 0.00853 |
| 3.85E-5 | 0.00889 |
| 3.78E-5 | 0.00951 |
| 3.7E-5 | 0.01018 |
| 3.65E-5 | 0.01097 |
| 3.56E-5 | 0.012 |
| 3.53E-5 | 0.01313 |
| 3.48E-5 | 0.01405 |
| 3.44E-5 | 0.0149 |
| 3.34E-5 | 0.01579 |
| 3.34E-5 | 0.0168 |
| 3.24E-5 | 0.01822 |
| 3.22E-5 | 0.01887 |
| 3.19E-5 | 0.0197 |
| 3.15E-5 | 0.02048 |
| 3.1E-5 | 0.02097 |
| 3.06E-5 | 0.02182 |
| 3.04E-5 | 0.0223 |
| 2.95E-5 | 0.02254 |
| 2.93E-5 | 0.02297 |
| 2.91E-5 | 0.02316 |
| 2.87E-5 | 0.02341 |
| 2.84E-5 | 0.02382 |
| 2.82E-5 | 0.02393 |
| 2.72E-5 | 0.02385 |
| 2.72E-5 | 0.02392 |
| 2.69E-5 | 0.02399 |
| 2.62E-5 | 0.02437 |
| 2.61E-5 | 0.02442 |
| 2.62E-5 | 0.02428 |
| 2.57E-5 | 0.02473 |
| 2.55E-5 | 0.02482 |
| 2.51E-5 | 0.02481 |
| 2.43E-5 | 0.02475 |
| 2.41E-5 | 0.02497 |
| 2.38E-5 | 0.02526 |
| 2.34E-5 | 0.02552 |
| 2.33E-5 | 0.02515 |
| 2.25E-5 | 0.02525 |
| 2.2E-5 | 0.02499 |
| 2.19E-5 | 0.02493 |
| 2.12E-5 | 0.02519 |
| 2.12E-5 | 0.02536 |
| 2.12E-5 | 0.02536 |
| 2.06E-5 | 0.02553 |
| 2.08E-5 | 0.02524 |
| 2.01E-5 | 0.02542 |
| 1.95E-5 | 0.02562 |
| 1.9E-5 | 0.02569 |
| 1.92E-5 | 0.02566 |
| 1.85E-5 | 0.02555 |
| 1.82E-5 | 0.02532 |
| 1.81E-5 | 0.0255 |
| 1.82E-5 | 0.02591 |
| 1.75E-5 | 0.02606 |
| 1.73E-5 | 0.02608 |
| 1.66E-5 | 0.02598 |
| 1.65E-5 | 0.02578 |
| 1.63E-5 | 0.02592 |
| 1.54E-5 | 0.02586 |
| 1.56E-5 | 0.02579 |
| 1.53E-5 | 0.02576 |
| 1.49E-5 | 0.02574 |
| 1.41E-5 | 0.02574 |
| 1.42E-5 | 0.02573 |
| 1.37E-5 | 0.02599 |
| 1.35E-5 | 0.02602 |
| 1.34E-5 | 0.02606 |
| 1.25E-5 | 0.02598 |
| 1.25E-5 | 0.02607 |
| 1.21E-5 | 0.02619 |
| 1.2E-5 | 0.02612 |
| 1.12E-5 | 0.02588 |
| 1.13E-5 | 0.02572 |
| 1.08E-5 | 0.02601 |
| 1.08E-5 | 0.02614 |
| 1.02E-5 | 0.02624 |
| 9.6E-6 | 0.02625 |
| 9.2E-6 | 0.02618 |
| 9.5E-6 | 0.0262 |
| 8.6E-6 | 0.02624 |
| 8.8E-6 | 0.02646 |
| 8.2E-6 | 0.02627 |
| 7.9E-6 | 0.02623 |
| 7.5E-6 | 0.02608 |
| 7.4E-6 | 0.02614 |
| 6.8E-6 | 0.02632 |
| 6.7E-6 | 0.02636 |
| 6.4E-6 | 0.02624 |
| 6E-6 | 0.02601 |
| 5.5E-6 | 0.02573 |
| 5E-6 | 0.0258 |
| 4.8E-6 | 0.02588 |
| 4.5E-6 | 0.02593 |
| 4.6E-6 | 0.02575 |
| 4.1E-6 | 0.02598 |
| 3.4E-6 | 0.0263 |
| 3.7E-6 | 0.02656 |
| 2.9E-6 | 0.02644 |
| 3E-6 | 0.02662 |
| 2.1E-6 | 0.0263 |
| 2.1E-6 | 0.02637 |
| 1.9E-6 | 0.02667 |
| 1.3E-6 | 0.02685 |
| 1.3E-6 | 0.02687 |
| 5E-7 | 0.02704 |
| 5E-7 | 0.02737 |
| 1.3E-6 | 0.02695 |
| 1.1E-6 | 0.02657 |
| 8E-7 | 0.02652 |
| 3E-7 | 0.02652 |
| 0 | 0.02699 |
| -6E-7 | 0.0272 |
| -1E-7 | 0.02712 |
| -9E-7 | 0.0269 |
| -1E-6 | 0.02657 |
| -1.4E-6 | 0.02649 |
| -1.9E-6 | 0.02641 |
| -2.2E-6 | 0.02654 |
| -2.5E-6 | 0.02614 |
| -2.6E-6 | 0.02613 |
| -3.4E-6 | 0.02604 |
| -3.3E-6 | 0.02605 |
| -3.8E-6 | 0.02618 |
| -3.8E-6 | 0.02603 |
| -4.2E-6 | 0.02582 |
| -4.7E-6 | 0.02597 |
| -4.6E-6 | 0.02579 |
| -5.1E-6 | 0.02599 |
| -5.7E-6 | 0.02592 |
| -6.1E-6 | 0.02617 |
| -6.3E-6 | 0.02617 |
| -7E-6 | 0.02634 |
| -6.9E-6 | 0.0263 |
| -7.5E-6 | 0.0263 |
| -7.7E-6 | 0.02623 |
| -8E-6 | 0.02598 |
| -8.4E-6 | 0.02604 |
| -8.5E-6 | 0.02612 |
| -9E-6 | 0.02623 |
| -9.2E-6 | 0.02612 |
| -9.9E-6 | 0.02611 |
| -9.9E-6 | 0.02622 |
| -1.02E-5 | 0.02634 |
| -1.11E-5 | 0.02638 |
| -1.07E-5 | 0.02619 |
| -1.16E-5 | 0.02619 |
| -1.12E-5 | 0.02619 |
| -1.23E-5 | 0.02599 |
| -1.24E-5 | 0.02609 |
| -1.26E-5 | 0.02627 |
| -1.29E-5 | 0.02608 |
| -1.3E-5 | 0.0262 |
| -1.38E-5 | 0.02618 |
| -1.35E-5 | 0.02598 |
| -1.43E-5 | 0.026 |
| -1.47E-5 | 0.02607 |
| -1.48E-5 | 0.02579 |
| -1.56E-5 | 0.02559 |
| -1.56E-5 | 0.02584 |
| -1.56E-5 | 0.0258 |
| -1.67E-5 | 0.02574 |
| -1.7E-5 | 0.02588 |
| -1.74E-5 | 0.02584 |
| -1.76E-5 | 0.02566 |
| -1.72E-5 | 0.02542 |
| -1.78E-5 | 0.02558 |
| -1.83E-5 | 0.02565 |
| -1.86E-5 | 0.0253 |
| -1.91E-5 | 0.02529 |
| -1.93E-5 | 0.0256 |
| -1.96E-5 | 0.02573 |
| -1.98E-5 | 0.02573 |
| -2.04E-5 | 0.02562 |
| -2.07E-5 | 0.02548 |
| -2.11E-5 | 0.02547 |
| -2.14E-5 | 0.02565 |
| -2.15E-5 | 0.02541 |
| -2.21E-5 | 0.02529 |
| -2.24E-5 | 0.02525 |
| -2.2E-5 | 0.02521 |
| -2.32E-5 | 0.02517 |
| -2.36E-5 | 0.02512 |
| -2.39E-5 | 0.02503 |
| -2.43E-5 | 0.02487 |
| -2.44E-5 | 0.02475 |
| -2.45E-5 | 0.02465 |
| -2.52E-5 | 0.02451 |
| -2.51E-5 | 0.02438 |
| -2.57E-5 | 0.02422 |
| -2.66E-5 | 0.02423 |
| -2.61E-5 | 0.02405 |
| -2.69E-5 | 0.02376 |
| -2.72E-5 | 0.02384 |
| -2.76E-5 | 0.02345 |
| -2.81E-5 | 0.02302 |
| -2.82E-5 | 0.0232 |
| -2.9E-5 | 0.02315 |
| -2.92E-5 | 0.02273 |
| -2.98E-5 | 0.02232 |
| -3.01E-5 | 0.0216 |
| -3.08E-5 | 0.021 |
| -3.03E-5 | 0.02052 |
| -3.14E-5 | 0.01978 |
| -3.2E-5 | 0.01894 |
| -3.22E-5 | 0.01813 |
| -3.35E-5 | 0.01707 |
| -3.32E-5 | 0.01612 |
| -3.44E-5 | 0.01517 |
| -3.5E-5 | 0.01384 |
| -3.52E-5 | 0.01314 |
| -3.65E-5 | 0.01238 |
| -3.75E-5 | 0.01131 |
| -3.84E-5 | 0.01052 |
| -3.84E-5 | 0.00982 |
| -4.06E-5 | 0.00922 |
| -4.08E-5 | 0.00878 |
| -4.1E-5 | 0.00841 |
| -4.36E-5 | 0.00811 |
| -4.35E-5 | 0.00781 |
| -4.52E-5 | 0.00759 |
| -4.65E-5 | 0.00743 |
| -4.79E-5 | 0.00735 |
| -4.81E-5 | 0.0073 |
| -5.07E-5 | 0.00733 |
| -5.1E-5 | 0.00745 |
| -5.12E-5 | 0.00754 |
| -5.36E-5 | 0.0077 |
| -5.34E-5 | 0.00784 |
| -5.51E-5 | 0.008 |
| -5.59E-5 | 0.0083 |
| -5.67E-5 | 0.00854 |
| -5.78E-5 | 0.00879 |
| -5.82E-5 | 0.00903 |
| -5.99E-5 | 0.00914 |
| -6.08E-5 | 0.00933 |
| -6.15E-5 | 0.00951 |
| -6.21E-5 | 0.00974 |
| -6.33E-5 | 0.01005 |
| -6.35E-5 | 0.01024 |
| -6.4E-5 | 0.0104 |
| -6.59E-5 | 0.01054 |
| -6.59E-5 | 0.01065 |
| -6.72E-5 | 0.01084 |
| -6.78E-5 | 0.01101 |
| -6.9E-5 | 0.01108 |
| -6.91E-5 | 0.01114 |
| -6.99E-5 | 0.01125 |
| -7.08E-5 | 0.01133 |
| -7.17E-5 | 0.01147 |
| -7.28E-5 | 0.01156 |
| -7.26E-5 | 0.01154 |
| -7.46E-5 | 0.0116 |
| -7.47E-5 | 0.01165 |
| -7.49E-5 | 0.01164 |
| -7.62E-5 | 0.01175 |
| -7.65E-5 | 0.01179 |
| -7.73E-5 | 0.01182 |
| -7.81E-5 | 0.01182 |
| -7.88E-5 | 0.01193 |
| -7.95E-5 | 0.01202 |
| -8.07E-5 | 0.01212 |
| -8.08E-5 | 0.01216 |
| -8.1E-5 | 0.01227 |
| -8.26E-5 | 0.01232 |
| -8.25E-5 | 0.01231 |
| -8.39E-5 | 0.0123 |
| -8.43E-5 | 0.01238 |
| -8.53E-5 | 0.01251 |
| -8.54E-5 | 0.0125 |
| -8.63E-5 | 0.01265 |
| -8.7E-5 | 0.01275 |
| -8.74E-5 | 0.01283 |
| -8.86E-5 | 0.0129 |
| -8.86E-5 | 0.01291 |
| -8.96E-5 | 0.01291 |
| -9.06E-5 | 0.01298 |
| -9.04E-5 | 0.01308 |
| -9.16E-5 | 0.01311 |
| -9.26E-5 | 0.0131 |
| -9.31E-5 | 0.01306 |
| -9.36E-5 | 0.0131 |
| -9.43E-5 | 0.01324 |
| -9.44E-5 | 0.01332 |
| -9.54E-5 | 0.01331 |
| -9.65E-5 | 0.0133 |
| -9.63E-5 | 0.01325 |
| -9.74E-5 | 0.01329 |
| -9.79E-5 | 0.01335 |
| -9.88E-5 | 0.01337 |
| -9.92E-5 | 0.01335 |
| -1E-4 | 0.01337 |
| -1.01E-4 | 0.0134 |
| -1.005E-4 | 0.01341 |
| -1.019E-4 | 0.01347 |
| -1.024E-4 | 0.01346 |
| -1.03E-4 | 0.01336 |
| -1.041E-4 | 0.01337 |
| -1.047E-4 | 0.01339 |
| -1.049E-4 | 0.01338 |
| -1.054E-4 | 0.01333 |
| -1.062E-4 | 0.01336 |
| -1.065E-4 | 0.0133 |
| -1.078E-4 | 0.01328 |
| -1.085E-4 | 0.01336 |
| -1.087E-4 | 0.01333 |
| -1.093E-4 | 0.01326 |
| -1.102E-4 | 0.01319 |
| -1.107E-4 | 0.01314 |
| -1.113E-4 | 0.01318 |
| -1.127E-4 | 0.0131 |
| -1.122E-4 | 0.013 |
| -1.139E-4 | 0.01291 |
| -1.139E-4 | 0.01288 |
| -1.147E-4 | 0.01274 |
| -1.153E-4 | 0.01263 |
| -1.161E-4 | 0.01262 |
| -1.162E-4 | 0.01252 |
| -1.174E-4 | 0.01234 |
| -1.186E-4 | 0.01225 |
| -1.189E-4 | 0.01212 |
| -1.197E-4 | 0.012 |
| -1.207E-4 | 0.0119 |
| -1.209E-4 | 0.01182 |
| -1.22E-4 | 0.01176 |
| -1.231E-4 | 0.0117 |
| -1.237E-4 | 0.01162 |
| -1.236E-4 | 0.0116 |
| -1.253E-4 | 0.01159 |
| -1.253E-4 | 0.0115 |
| -1.26E-4 | 0.01148 |
| -1.276E-4 | 0.01146 |
| -1.28E-4 | 0.01145 |
| -1.287E-4 | 0.01143 |
| -1.294E-4 | 0.01144 |
| -1.304E-4 | 0.01153 |
| -1.306E-4 | 0.01155 |
| -1.313E-4 | 0.01157 |
| -1.326E-4 | 0.01161 |
| -1.331E-4 | 0.01164 |
| -1.338E-4 | 0.01171 |
| -1.341E-4 | 0.01168 |
| -1.349E-4 | 0.01168 |
| -1.356E-4 | 0.01175 |
| -1.363E-4 | 0.01179 |
| -1.375E-4 | 0.01179 |
| -1.381E-4 | 0.01188 |
| -1.39E-4 | 0.01191 |
| -1.39E-4 | 0.01188 |
| -1.403E-4 | 0.01193 |
| -1.409E-4 | 0.01197 |
| -1.42E-4 | 0.01196 |
| -1.423E-4 | 0.01198 |
| -1.425E-4 | 0.01199 |
| -1.439E-4 | 0.01196 |
| -1.439E-4 | 0.01195 |
| -1.458E-4 | 0.01196 |
| -1.459E-4 | 0.01205 |
| -1.466E-4 | 0.01206 |
| -1.47E-4 | 0.01205 |
| -1.48E-4 | 0.01204 |
| -1.483E-4 | 0.012 |
| -1.494E-4 | 0.01202 |
| -1.504E-4 | 0.01197 |
| -1.505E-4 | 0.01207 |
| -1.516E-4 | 0.01203 |
| -1.527E-4 | 0.01205 |
| -1.523E-4 | 0.01213 |
| -1.533E-4 | 0.01213 |
| -1.543E-4 | 0.01218 |
| -1.55E-4 | 0.01215 |
| -1.557E-4 | 0.0122 |
| -1.565E-4 | 0.01224 |
| -1.57E-4 | 0.01214 |
| -1.584E-4 | 0.01211 |
| -1.588E-4 | 0.01216 |
| -1.588E-4 | 0.0121 |
| -1.597E-4 | 0.01214 |
| -1.609E-4 | 0.01219 |
| -1.612E-4 | 0.01214 |
| -1.617E-4 | 0.01214 |
| -1.63E-4 | 0.01216 |
| -1.632E-4 | 0.01216 |
| -1.638E-4 | 0.01218 |
| -1.651E-4 | 0.01221 |
| -1.652E-4 | 0.01221 |
| -1.659E-4 | 0.01222 |
| -1.667E-4 | 0.01227 |
| -1.678E-4 | 0.01221 |
| -1.682E-4 | 0.01223 |
| -1.689E-4 | 0.01228 |
| -1.692E-4 | 0.01229 |
| -1.701E-4 | 0.01219 |
| -1.71E-4 | 0.01222 |
| -1.714E-4 | 0.01224 |
| -1.726E-4 | 0.01222 |
| -1.731E-4 | 0.01219 |
| -1.732E-4 | 0.0122 |
| -1.744E-4 | 0.01225 |
| -1.748E-4 | 0.01217 |
| -1.757E-4 | 0.01216 |
| -1.767E-4 | 0.01223 |
| -1.773E-4 | 0.01222 |
| -1.78E-4 | 0.01222 |
| -1.788E-4 | 0.0123 |
| -1.789E-4 | 0.01223 |
| -1.798E-4 | 0.01218 |

**Data fitting code**
(Please contact Rafael Haenel rafaelhaenel@phas.ubc.ca for more information)

**Readme.txt**
First execute "0_data_processing.py". This takes the experimentally measured dIdV curve in the folder "raw_data" and saves supercurrent IV curves in the folder "processed_data"

Then run Mathematica notebook "1_fitting.nb" to perform the fitting and to generate all figures in the folder "figures"

**0_data_processing.py**
```
import numpy as np
import matplotlib.pyplot as plt
from scipy.optimize import curve_fit
from scipy.integrate import cumtrapz
import glob
import os
```

```python
import re
from scipy import interpolate
import scipy.optimize

plt.ion()
plt.close('all')

# load all file paths. Filenames are in the format dIdV[temp]mK.DAT and
# contain dIdV curves for temperature [temp]
files_unsorted = glob.glob('raw_data/*.DAT')
N = len(files_unsorted)

# extract all measured temperatures from filennames
temp = np.array([int(re.findall('\d+', f.replace('raw_data', ''))[0])
                 for f in files_unsorted])
idx = temp.argsort()
temp = temp[idx]  # mK
np.savetxt('processed_data/t_dependence/temp.csv', temp)

files = [files_unsorted[i] for i in idx]

def load():
    """
    this function loads all experimental dIdV curves into memory
    and returns np.arrays V (uV), dIdV (1/ohm) where V[:,i], dIdV[:,i] correspond
    to a dIdV curve measured at temperature temp[i]
    """
    L = 2500
    V = np.zeros((L, N))
    dIdV = np.zeros((L, N))

    # load all files
    for i in np.arange(N):
        v, didv = np.loadtxt(files[i], skiprows=1, unpack=True)
        V[0:len(v), i] = v * 1e6  # convert to uV
        dIdV[0:len(didv), i] = didv  # 1/ohm

    return V, dIdV

def y_intercept(x, y):
    """
    this function returns the y-intercept of the function y=f(x)
    defined as the interpolation of the function given by arrays x, y
    """
```

```python
    m1 = x*1
    m1[x > 0] = -10
    m2 = x*1
    m2[x < 0] = 10

    i = m1.argmax()
    j = m2.argmin()
    # return the interpolated value where curve crosses y-axis
    return y[i] + (y[j] - y[i])/(x[j]-x[i]) * (-x[i])

def center(dx, x, y):
    """
    moves the function defined by arrays x, y along the x-axis by dx
    then moves the function along y-axis such that y-intercept is zero
    then returns integral of (f(x) - (-f(-x)) )**2 from -10 to 10
    """
    yy = y - y_intercept(x-dx, y)

    f1 = interpolate.interp1d(x-dx, yy)
    f2 = interpolate.interp1d(-(x-dx), -yy)
    xx = np.linspace(-10, 10, 100)

    return np.sum((f1(xx) - f2(xx))**2)

def plot_dIdV(ind):
    """
    Plot a single dIdV curve with temperature temp[ind]
    """
    x = V[:, ind]
    y = dIdV[:, ind]
    T = temp[ind]
    cond = np.logical_and(x != 0, y != 0)
    x = x[cond]
    y = y[cond]

    idx = x.argsort()
    x = x[idx]
    y = y[idx]

    plt.plot(x, y, label='T='+str(T)+'mK')
    plt.ylabel('Diff. conductance dI/dV in $\Omega^{-1}$')
    plt.xlabel('Bias in $\mu$V')
    plt.title('T='+str(T)+'mK')
    plt.tight_layout()
```

```python
def linear_fit(ind, vmax=40, subtract=False, mirror=False, plot=False):
    T = temp[ind]  # mK
    v = V[:, ind]  # muV
    didv = dIdV[:, ind]  # 1/ohm

    # remove trailing zeroes in data array
    cond = np.logical_and(v != 0, didv != 0)
    v = v[cond]
    didv = didv[cond]

    # make sure data is sorted
    idx = v.argsort()
    v = v[idx]  # muV
    didv = didv[idx]  # 1/ohm = A/V

    # integrate data
    i = cumtrapz(didv, v)  # 1/ohm*muV = uA
    v = v[1:]

    # find the actual point of zero bias and zero current,
    # i.e. slightly move the I_V curve such that the particle hole condition
    # I(v) = -I(-V) is satisfied
    res = scipy.optimize.minimize(center, x0=0, args=(v, i))
    v -= res.x  # correct x-position
    i = i - y_intercept(v, i)  # correct y-position

    # print the value of x-correction to make sure that
    # this correction is in fact very small
    print(f'x-position of I-V curve has been adjusted by {res.x} uV for {T}mK')

    i = i[np.abs(v) < vmax]
    v = v[np.abs(v) < vmax]

    def f_lin(x, a):
        return a*x

    # find linear contribution that corresponds to normal current
    coeff, cov = curve_fit(f_lin, v, i)

    if subtract:
        # subtract normal current
        i -= f_lin(v, coeff)

    if plot:
```

```python
    plt.figure(figsize=(4, 3.6))
    plt.plot(v, i)
    if mirror:
        plt.plot(-v, -i, '-.')
    xfit = np.linspace(np.min(v), np.max(v), 1000)
    plt.plot(xfit, coeff*xfit, 'r--')
    plt.ylabel('Current I in $\mu A$')
    plt.xlabel('Bias in $\mu$V')
    plt.tight_layout()

    return v, i

def fit_coeffs(vmax=40):
    fit_res = np.zeros(len(temp))
    for i in np.arange(len(temp)):
        fit_res[i] = linear_fit(i, vmax=vmax)
    plt.figure(figsize=(4, 3.6))
    plt.plot(temp, fit_res, '.-')
    plt.xlabel('T in mK')
    plt.ylabel('Plateau conductance from lin. fit in $\Omega^{-1}$')
    plt.tight_layout()

def  plot_supercurrent(ind,  max=None, cutoff=10, subtract=True, save=False, mirror=False,
fitvmax=20):
    """
    This function performs the following operations
    on a dIdV curve recorded at temperature temp[i]:
    (1) integrate to obtain I-V curve
    (2) offset the data slightly such that particle hole symmetry
        I(V) = -I(-V) is respected, i.e. finding the true zero bias
        and zero current
    (3) subtract the linear contribution that corresponds to the
        normal current channel
    """
    T = temp[ind]
    v, i = linear_fit(ind, subtract=subtract, vmax=fitvmax)

    plt.plot(v, i, label='T='+str(T)+'mK')
    if mirror:
        plt.plot(-v, -i, label='$-I(-V)$')
    plt.xlim((-cutoff, cutoff))

    plt.tight_layout()
```

```python
    if ind != -1 and save:
        np.savetxt('processed_data/t_dependence/'+str(ind)+'x.csv', v)
        np.savetxt('processed_data/t_dependence/'+str(ind)+'y.csv', i)

    return v, i

# load data
V, dIdV = load()

# (Figure 3a) plot an exemplary I-V curve with a linear fit.
# The linear contribution corresponds to the normal part of the current
# that is subtracted
linear_fit(0, vmax=25, plot=True)
plt.savefig('figures/figure_3_a.pdf')

# (Figure 3b) Extract supercurrents, plot, and save data to
# folder "processed data"
plt.figure(figsize=(4, 3.6))
for i in np.arange(N):
    plot_supercurrent(i, subtract=True, save=True, fitvmax=20)
plt.ylim((-0.038, 0.038))
plt.xlim((-10, 10))
plt.xlabel('Bias in $\mu$V')
plt.ylabel('Supercurrent in $\mu$A')
plt.tight_layout()
plt.savefig('figures/figure_3_b.pdf')

def fraunhofer(b):
    # b in mT
    # Fraunhofer pattern of supercurrent
    phi0 = 2.067833848e-15  # Wb, flux quantum
    area = 4.13e-12  # m2, approximate area of the junction
    x = np.pi * b/1000 * area / phi0
    return np.abs(np.sin(x)/x)

def b_field_dependence():
    """
    This function takes the experimental data for B=0T, T=21mK
    and extrapolates it to finite magnetic fields by
    assuming a Fraunhofer-pattern B-dependence.
    The output data is saved to the folder "processed_data/b_dependece"
    for plotting with the Mathematica script "1_fitting.nb"
    """
```

```python
plt.figure(figsize=(4, 3.6))
# integrate dIdV with and without subtracting the supercurrent
x0, normal_and_super = plot_supercurrent(0, subtract=False, fitvmax=20)
_, supercurrent = plot_supercurrent(0, subtract=True, fitvmax=20)

# only keep data within +=10uV
cutoff = 10  # uV
cut = np.abs(x0) <= cutoff
x0 = x0[cut]
normal_and_super = normal_and_super[cut]
supercurrent = supercurrent[cut]

# clean up data
# remove datapoints that are recorded for the same bias
diffx0 = np.diff(x0)
x0 = x0[:-1][diffx0 != 0]
normal_and_super = normal_and_super[:-1][diffx0 != 0]
supercurrent = supercurrent[:-1][diffx0 != 0]

# ensure that supercurrent goes to 0 at +=10uV
# i.e. smooth out
th = 15
delta = 0.1
smooth_left = (np.tanh(delta*(x0+th))+1)/2
smooth_right = (-np.tanh(delta*(x0-th))+1)/2
supercurrent = supercurrent*smooth_left*smooth_right

# isolate the normal current
normal_current = normal_and_super - supercurrent

res = 300
bval = np.linspace(-1.5, 1.5, res)  # magnetic fields
# array that contains B-field dependence of supercurrent
fr = fraunhofer(bval)

# arrange in grid with bias axis and B-field axis
fr2 = np.tile(fr, (len(x0), 1)).T
red2 = np.tile(normal_current, (res, 1))
sc2 = np.tile(supercurrent, (res, 1))

# change the supercurrent according to its B-dependence
# then add normal current and modified supercurrent
full = red2 + sc2 * fr2

# compute the derivative to get differential conductance
cond = np.gradient(full, x0, axis=1)
```

```python
        cond = cond/np.max(cond)*4

        plt.figure()
        plt.pcolor(x0, bval, cond)
        plt.colorbar()

        # save data for plotting in Mathematica
        bfolder = 'processed_data/b_dependence/'
        np.savetxt(bfolder + 'data.txt', cond)
        np.savetxt(bfolder + 'bval.txt', np.tile(bval, (len(x0), 1)).T)
        np.savetxt(bfolder + 'V.txt', np.tile(x0, (len(bval), 1)))
        filt = np.abs(V[:, 0]) <= 10
        np.savetxt(bfolder + 'x.txt', V[filt, 0])
        np.savetxt(bfolder + 'y.txt', dIdV[filt, 0])

# (Figure 4c) fraunhofer pattern
b_field_dependence()
```

## Function definitions

```mathematica
Clear["Global`*"]
s = {15, FontFamily -> "Times New Roman"};
```

```mathematica
k = 8.617330 × 10^-5;(*eV/K*)
Δ = 50 × 10^-6; (*eV*)
hbar = 6.582119569 × 10^-16;(*eV s*)
ele = 1.602176634 × 10^-19; (*elementary charge in C*)
α = hbar / (k 2 ele 10^6);
Tc = 0.47;(*K*)
linspace[start_, stop_, n_: 100] := N[Table[x, {x, start, stop, (stop - start) / (n - 1)}]]

gap[T_] := 1 - Sqrt[2 Pi k T / Δ] Exp[-Δ / (k T)]
(*temperature dependence of BCS gap (first order expansion at small T)*)
gap2[T_] := Tanh[1.74 Sqrt[Tc / T - 1]](* alternatively, 
use this phenomenological temperature dependence*)

(*fitting function f=I(V) function according to Eq. (1). The input arguments are
   Bias v in μV
   Critical supercurrent in μA
   R1,R2 in Ω
   T in K *)
f[v_, i0_, r1_, r2_, T_] := 
 N[i0 gap[T] Im[BesselI[1 - I α v / ((r1 + r2) T), α i0 gap[T] / T] / 
    BesselI[-I α v / ((r1 + r2) T), α i0 gap[T] / T]]]
(*version fitting function that takes array of
 voltages vList instead of single quantity*)
fpoints[vlist_, i0_?NumericQ, r1_?NumericQ, r2_?NumericQ, T_?NumericQ] :=
 N[Table[f[v, i0, r1, r2, T], {v, vlist}]]
vmax = 5;
```

## Load data sets for a selected number of temperatures.

```
In[ ]:= dir = NotebookDirectory[] <> "processed_data\\t_dependence\\";
idx = {3, 5, 7, 9, 11, 13, 14};
temp = IntegerPart@(Identity@@@ Import[dir <> "temp.csv"]);
xar = {}; yar = {};
plotsd = {};
For[ind = 1, ind ≤ Length[idx], ind++,
 path = dir <> ToString[idx[[ind]] - 1];
 Print[path <> "x.csv"];
 x = Identity@@@ Import[path <> "x.csv"];
 y = Identity@@@ Import[path <> "y.csv"];
 xy = Transpose@{x, y};
 xy = Select[xy, 0 <= #[[1]] ≤ vmax &];
 x = N@Transpose[xy][[1]];
 y = N@Transpose[xy][[2]];
 AppendTo[xar, x]; AppendTo[yar, y];
 AppendTo[plotsd, ListPlot[xy, PlotRange → {0, 0.03},
   Frame → True, PlotLabel → "T=" <> ToString[temp[[ind]]] <> "mK",
   FrameLabel → {Style["V in μV", s], Style["I in μA", s]}]];
]
plotsd
```

C:\Users\rafae\Desktop\ZBCP LJ16617
   Yu\fig_3_temperature_dependence\processed_data\t_dependence\2x.csv

C:\Users\rafae\Desktop\ZBCP LJ16617
   Yu\fig_3_temperature_dependence\processed_data\t_dependence\4x.csv

C:\Users\rafae\Desktop\ZBCP LJ16617
   Yu\fig_3_temperature_dependence\processed_data\t_dependence\6x.csv

C:\Users\rafae\Desktop\ZBCP LJ16617
   Yu\fig_3_temperature_dependence\processed_data\t_dependence\8x.csv

C:\Users\rafae\Desktop\ZBCP LJ16617
   Yu\fig_3_temperature_dependence\processed_data\t_dependence\10x.csv

C:\Users\rafae\Desktop\ZBCP LJ16617
   Yu\fig_3_temperature_dependence\processed_data\t_dependence\12x.csv

C:\Users\rafae\Desktop\ZBCP LJ16617
   Yu\fig_3_temperature_dependence\processed_data\t_dependence\13x.csv

Out[ ]= {
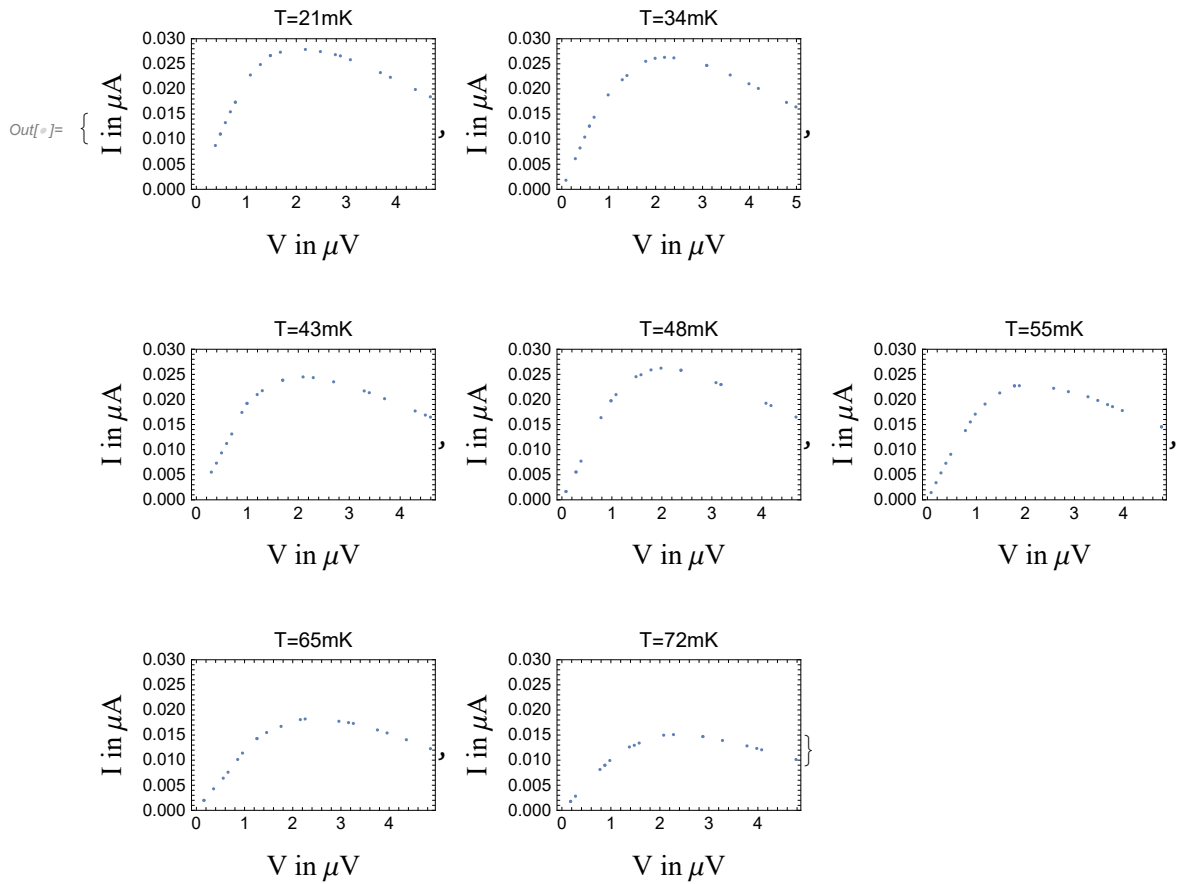
}

## Define optimization function and minimize

In[ ]:= (* define least square optimization function
 that fits all of above datasets at the same time. *)
mf[i0_?NumericQ, r1_?NumericQ, r2_?NumericQ] :=
 N[ $\sum_{i=1}^{\text{Length}[\text{idx}]}$ (Norm[ yar[[i]] - fpoints[xar[[i]] + r2 yar[[i]],
     i0, r1, r2, $10^{-3}$ temp[[idx[[i]]]] ] ]^2 ) ]
(*minimize and return fit parameters, i0 (μA), r1,r2 (Ω)*)
fp = FindMinimum[{mf[i0, r1, r2], i0 > 0.02, i0 < 0.04},
  {{i0, 0.0348}, {r1, 58.6}, {r2, 88.4}}, AccuracyGoal → 3]

Out[ ]= {0.000252192, {i0 → 0.0348167, r1 → 59.3078, r2 → 91.78}}

In[ ]:= **Plot fits**

Out[ ]= fits Plot

```
In[ ]:= plots = {};
    For[sel = 1, sel ≤ Length[idx], sel++,
     T = temp[[idx[[sel]] ]] 10^-3;
     xvals = linspace[0, vmax * 2, 100];
     res = fpoints[xvals , i0, r1, r2, T] /. fp[[2]];
     vvals = xvals - r2 * res /. fp[[2]];
     xysel = Transpose@{xar[[sel]], N[yar[[sel]]]};
     pl = Show[{
         ListPlot[Transpose@{vvals, res},
          PlotRange → {{0, vmax}, {0, i0 /. fp[[2]]}}, Joined → True,
          Frame → True, FrameLabel → {Style["V in μV", s], Style["I in μA", s]},
          PlotLegends → Placed[{"Theory"}, {Scaled[{0.85, 0.9}], {0.5, 0.5}}],
          PlotStyle → ColorData[6, "ColorList"],
          AspectRatio → 1,
          FrameTicksStyle → Directive[s],
          FrameStyle → Thickness[0.002],
          ImageSize → Medium,
          Epilog →
           {Text[Style["T=" <> ToString[temp[[idx[[sel]]]]] <> "mK", s], Scaled[{0.16, 0.9}]]}
         ],

         ListPlot[xysel, PlotLegends → Placed[{"Exp."}, {Scaled[{0.85, 0.8}], {0.5, 0.5}}],
          PlotStyle → ColorData[6, "ColorList"]]
        }];
     AppendTo[plots, pl];
    ]
    plots
```

Out[ ]= { 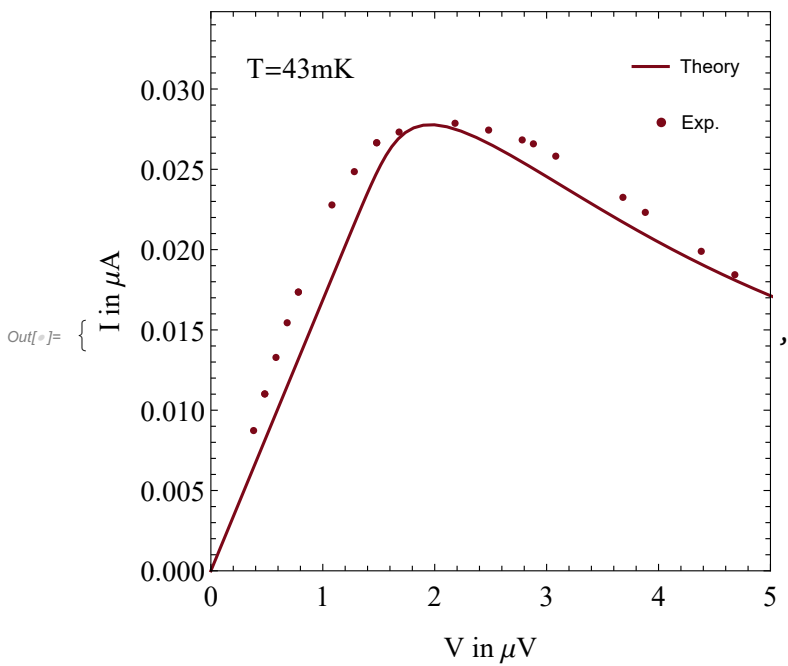 ,

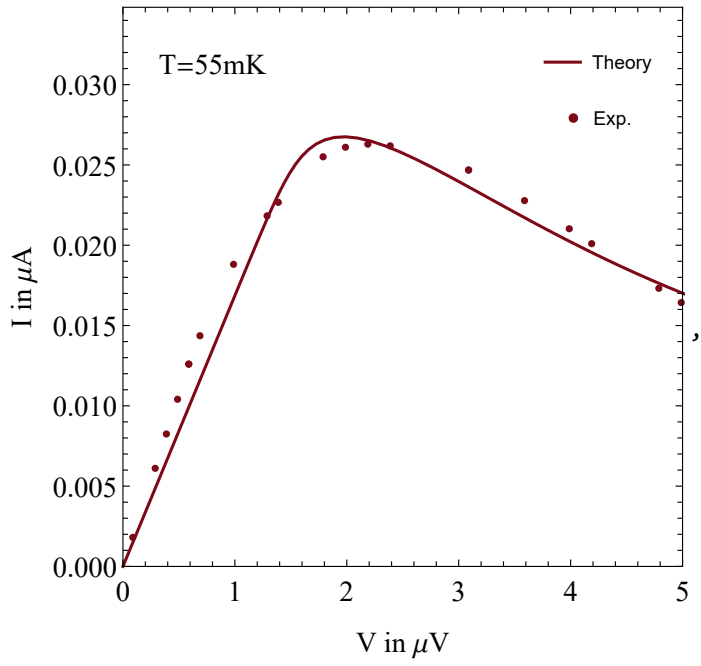,

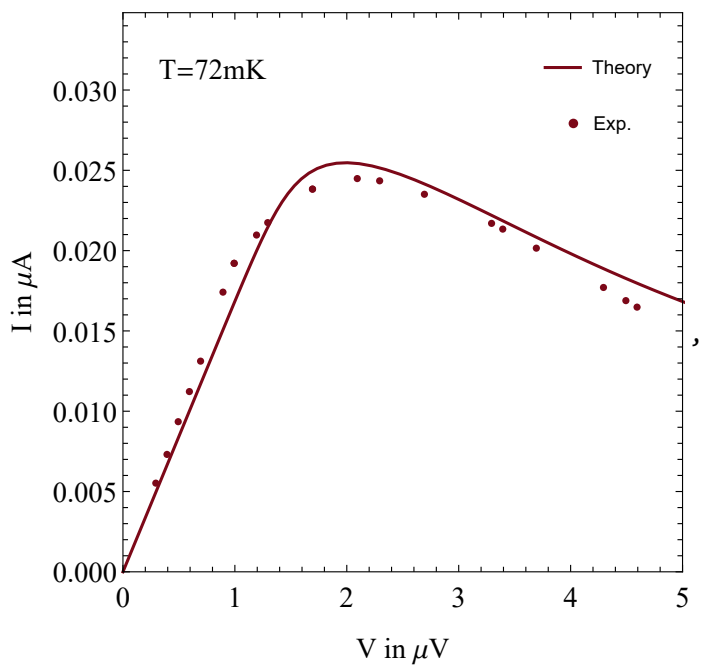,

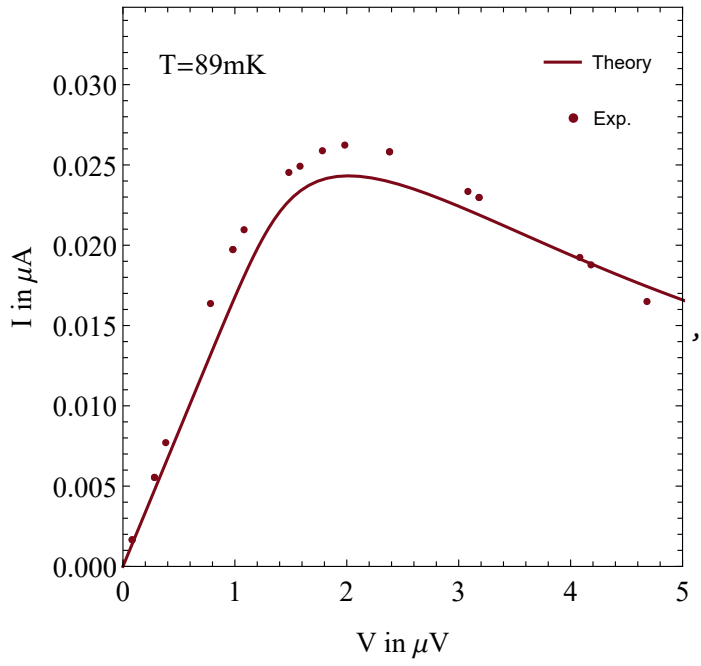

,

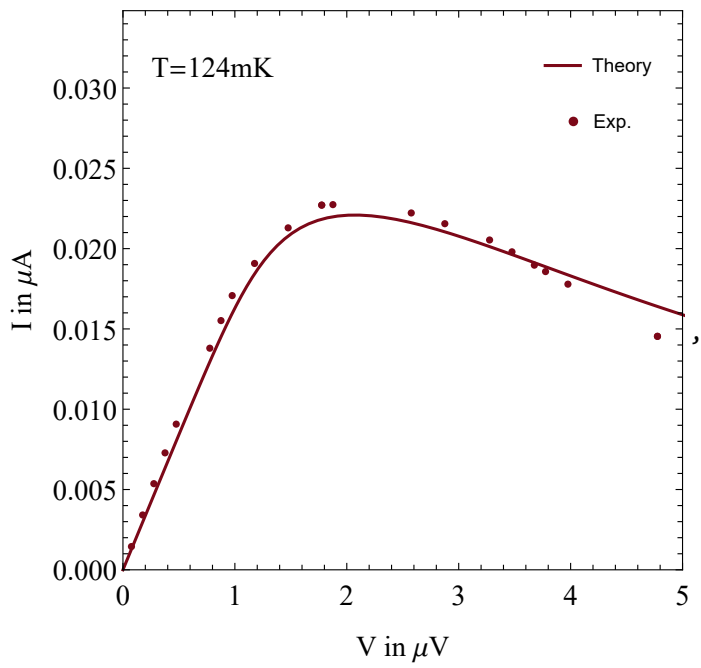

,

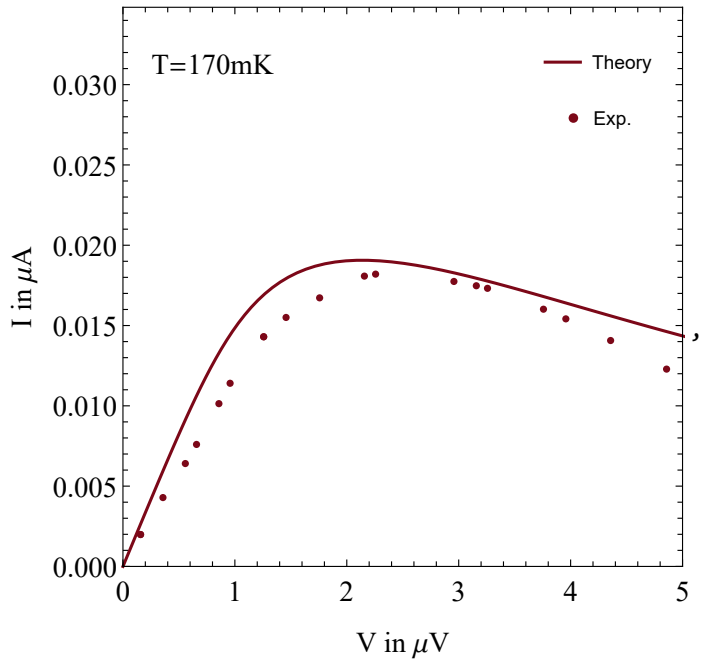

,

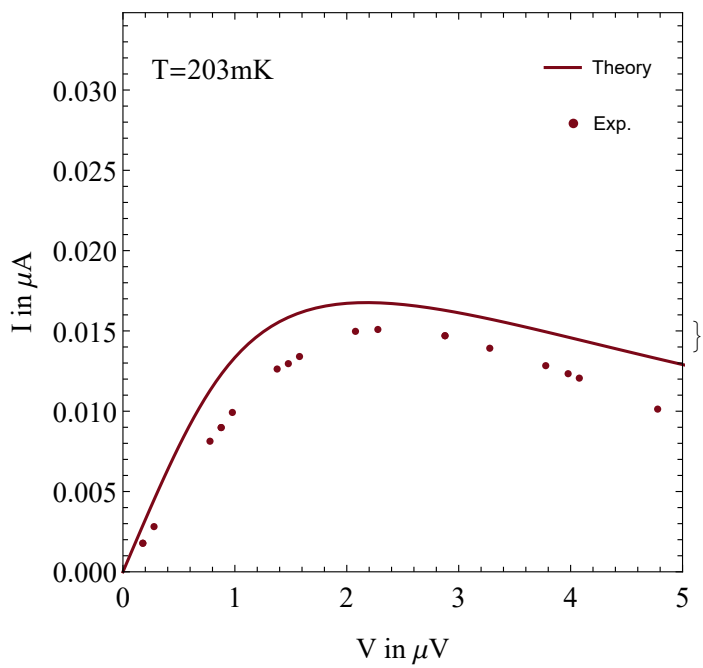

}

Save fits for all temperatures in folder "figures/fits"

```mathematica
folder = NotebookDirectory[] <> "figures\\fits\\";
Imaxfit = {};
Imaxexp = {};
For[i = 1, i ≤ Length[temp], i++,
 x = Identity @@@ Import[dir <> ToString[i - 1] <> "x.csv"];
 y = Identity @@@ Import[dir <> ToString[i - 1] <> "y.csv"];
 
 xy = Transpose@{x, y};
 xy = Select[xy, 0 < #[[1]] <= 5 &];
 x = N@Transpose[xy][[1]];
 y = N@Transpose[xy][[2]];
 
 xysel = Transpose@{x, N[y * 1000]};
 
 T = temp[[i]] 10^-3;
 
 xvals = linspace[0, vmax * 2, 100];
 res = fpoints[xvals , i0, r1, r2, T] /. fp[[2]];
 vvals = xvals - r2 * res /. fp[[2]];
 AppendTo[Imaxfit, Max[res]]; (*get maximum current of fit*)
 AppendTo[Imaxexp, Max [N[y]]];
 (*get maximum current of experimental data*)
 Export[folder <> "res" <> ToString[i - 1] <> ".pdf",
  
  Show[{
    ListPlot[Transpose@{vvals, res * 1000},
     PlotRange → {{0, vmax}, {0, i0 * 1000 /. fp[[2]]}}, Joined → True,
     Frame → True, FrameLabel → {Style["V in μV", s], Style["I in nA", s]},
     PlotLegends → Placed[{"Theory"}, {Scaled[{0.85, 0.9}], {0.5, 0.5}}],
     PlotStyle → ColorData[6, "ColorList"],
     AspectRatio → 1,
     FrameTicksStyle → Directive[s],
     FrameStyle → Thickness[0.002],
     ImageSize → Medium,
     Epilog → {Text[Style["T=" <> ToString[T 10^3] <> "mK", s], Scaled[{0.16, 0.9}]]}
     ],
    
    ListPlot[xysel, PlotLegends → Placed[{"Exp."}, {Scaled[{0.85, 0.8}], {0.5, 0.5}}],
     PlotStyle → ColorData[6, "ColorList"]]
    }]
  
  ]
 ]
```

## Plot and save maximum current vs. temperature

and current maximum Plot save vs. temperature

```
In[ ]:= mc = Show[{ListPlot[Transpose@{temp, Imaxfit * 1000}, Joined → True, Frame → True,
       FrameLabel → {"Temperature in mK", "Maximum current in nA"}, PlotLegends → {"fit"}],
      ListPlot[Transpose@{temp, Imaxexp * 1000}, Joined → False, PlotLegends → {"exp"}]},
     LabelStyle → Directive[FontSize → 15, Bold]]
    Export[folder <> "..//figure_3_f.pdf", mc]
```

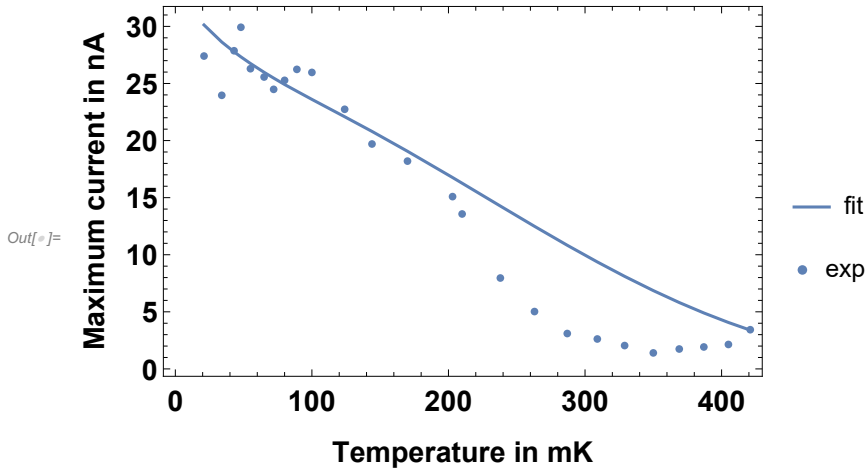

```
Out[ ]= C:\Users\rafae\Desktop\ZBCP LJ16617
       Yu\fig_3_temperature_dependence\figures\fits\..//figure_3_f.pdf
```

## Fraunhofer Pattern

```
In[ ]:= ClearAll[MPLColorMap]
    << "http://pastebin.com/raw/pFsb4ZBS";
    bfolder = NotebookDirectory[] <> "//processed_data//b_dependence";
    data = Import[bfolder <> "//data.txt", "Table"];
    bval = Import[bfolder <> "//bval.txt", "Table"];
    vval = Import[bfolder <> "//V.txt", "Table"];

    xval = Flatten@Import[bfolder <> "//x.txt", "Table"];
    yval = Flatten@Import[bfolder <> "//y.txt", "Table"];
    pl = ArrayFlatten[Transpose[{bval, vval, data}, {3, 2, 1}], 1];

    bdep = ListDensityPlot[pl,
      ColorFunction → (ColorData["Rainbow"][Rescale[#, {1.35, 3}]] &), PlotRange → All,
      Axes → False, ColorFunctionScaling → False,
      FrameLabel → {Style["B (mT)", s], Style["V_dc (μV)", s]},
      AspectRatio → 0.83,
      FrameTicksStyle → Directive[s],
      FrameStyle → Thickness[0.002], PlotLegends → Automatic, LabelStyle → Directive[15]]
    Export[folder <> "..//figure_4_c.pdf", bdep]
```

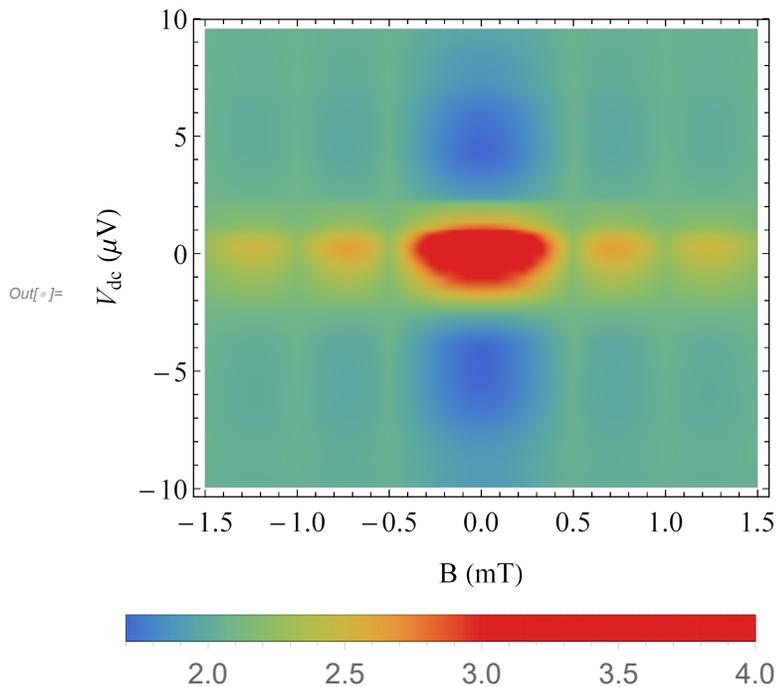

```
Out[ ]= C:\Users\rafae\Desktop\ZBCP LJ16617
    Yu\fig_3_temperature_dependence\figures\fits\..//figure_4_c.pdf
```